\documentclass[sigconf,nonacm]{acmart}

\usepackage{algorithm,algpseudocode,setspace,subcaption,multirow}
\usepackage{xcolor}
\AtBeginDocument{%
  \providecommand\BibTeX{{%
    \normalfont B\kern-0.5em{\scshape i\kern-0.25em b}\kern-0.8em\TeX}}}

\copyrightyear{2022}
\acmYear{2022}
\setcopyright{acmlicensed}\acmConference[HPC Asia2022]{International Conference on High Performance Computing in Asia-Pacific Region}{January 12--14, 2022}{Virtual Event, Japan}
\acmBooktitle{International Conference on High Performance Computing in Asia-Pacific Region (HPC Asia2022), January 12--14, 2022, Virtual Event, Japan}
\acmPrice{15.00}
\acmDOI{10.1145/3492805.3492808}
\acmISBN{978-1-4503-8498-8/22/01}

\begin{document}


\title[A High-Fidelity Flow Solver for Unstructured Meshes on Field-Programmable Gate Arrays]{\vspace{-20pt}A High-Fidelity Flow Solver for Unstructured Meshes on Field-Programmable Gate Arrays: \textit{Design, Evaluation, and Future Challenges}}


\author{Martin Karp}
\affiliation{
  \institution{KTH Royal Institute of Technology}
  \city{Stockholm}
  \country{Sweden}
}
\email{makarp@kth.se}

\author{Artur Podobas}
\affiliation{
  \institution{KTH Royal Institute of Technology}
  \city{Stockholm}
  \country{Sweden}
}
\email{podobas@kth.se}

\author{Tobias Kenter}
\affiliation{
  \institution{Paderborn University}
  \city{Paderborn}
  \country{Germany}
}
\email{kenter@uni-paderborn.de}

\author{Niclas Jansson}
\affiliation{
  \institution{KTH Royal Institute of Technology}
  \city{Stockholm}
  \country{Sweden}
}
\email{njansson@kth.se}

\author{Christian Plessl}
\affiliation{
  \institution{Paderborn University}
  \city{Paderborn}
  \country{Germany}
}
\email{christian.plessl@upb.de}

\author{Philipp Schlatter}
\affiliation{
  \institution{KTH Royal Institute of Technology}
  \city{Stockholm}
  \country{Sweden}
}
\email{pschlatt@mech.kth.se}

\author{Stefano Markidis}
\affiliation{
  \institution{KTH Royal Institute of Technology}
  \city{Stockholm}
  \country{Sweden}
}
\email{markidis@kth.se}

\begin{abstract}
The impending termination of Moore's law motivates the search for new forms of computing to continue the performance scaling we have grown accustomed to. Among the many emerging \textit{Post-Moore} computing candidates, perhaps none is as salient as the Field-Programmable Gate Array (FPGA), which offers the means of specializing and customizing the hardware to the computation at hand. 

In this work, we design a custom FPGA-based accelerator for a computational fluid dynamics (CFD) code. Unlike prior work -- which often focuses on accelerating small kernels -- we target the entire Poisson solver on unstructured meshes based on the high-fidelity spectral element method (SEM) used in modern state-of-the-art CFD systems. We model our accelerator using an analytical performance model based on the I/O cost of the algorithm. We empirically evaluate our accelerator on a state-of-the-art Intel Stratix 10 FPGA in terms of performance and power consumption and contrast it against existing solutions on general-purpose processors (CPUs). Finally, we propose a data movement-reducing technique where we compute geometric factors on the fly, which yields significant (700+ Gflop/s) single-precision performance and an upwards of 2x reduction in runtime for the local evaluation of the Laplace operator.

We end the paper by discussing the challenges and opportunities of using reconfigurable architecture in the future, particularly in the light of emerging (not yet available) technologies.
\end{abstract}

\ccsdesc[500]{Hardware~Hardware accelerators}
\ccsdesc[500]{Computer systems organization~Reconfigurable computing}
\ccsdesc[300]{Mathematics of computing~Solvers}
\ccsdesc[100]{Computing methodologies~Massively parallel and high-performance simulations}


\keywords{Spectral element method, Field-programmable gate array, Conjugate gradient method, High-level synthesis}

\maketitle
\renewcommand{\shortauthors}{Karp et al.}
\section{Introduction}

The end of Dennard's scaling~\cite{jouppi2018motivation} and the impending termination of Moore's law~\cite{waldrop2016chips} is today forcing researchers to actively search for alternative hardware solutions. Several new and intrusive so-called \textit{Post-Moore} technologies are emerging, including, for example, neuromorphic- or quantum-computing~\cite{schuman2017survey,gyongyosi2019survey}. Unfortunately, many of these are either very niched (e.g., neuromorphic computing) or nonviable for the foreseeable future (e.g., quantum computing). There is, however, one type of post-Moore technology~\cite{vetter2017architectures} that preserves the most salient properties of the general-purpose computers that we have grown used to, while at the same time offer remedies to overcome the performance- and memory-bottlenecks of the former: the reconfigurable architecture.

Reconfigurable architectures, such as Field-Programmable Gate Arrays (FPGAs) or Coarse-Grained Reconfigurable Architectures (CGRAs)~\cite{podobas2020survey}, are systems that aspire to provide some form of silicon plasticity and can thus counter the end of Moore's law-- we do not need more transistors; instead, we need to re-purpose the ones we have and specialize them to the application at hand. Reconfigurable systems can also overcome the expensive von Neumann bottleneck (which can incur a near three-orders-magnitude energy consumption overhead inside a CPU~\cite{jouppi2018motivation}) by spatially map the computation without the need for instruction decoding or unnecessary (register-to-register) data transfers.

Today, a large fraction of scientific applications are bound by data movement\cite{ivanov2021data,marjanovic2014performance}. Data movement has (historically) been more expensive for several reasons: \textbf{(i)} external memory bandwidth is growing at a slower pace compared to computation, and \textbf{(ii)} the sheer amount of data is increasing in volume. At the same time, emerging technologies such as 3D stacking~\cite{zhang2014survey} can be expected to alleviate the impact of data movement. To remedy the cost of data movement, there is thus a dire need to shift towards a more data movement-centric view on optimizations e.g. \cite{kwasniewski2019red,Kwasniewski_2021}, and it is imperative to model, understand, and optimize for the data movement bounds in which target applications operate. Among the many application domains that suffer from the consequences of said imbalance in data movement capabilities is computational fluid dynamics (CFD)~\cite{slotnick2014cfd}.  

This work investigates the opportunities and challenges for accelerating CFD using modern reconfigurable architectures and honoring, in particular, the data movement aspects of said computation. We focus on the spectral element method (SEM)~\cite{deville2002high}, which is a high-order  method used in  state-of-the-art, high-fidelity CFD solvers such as Nek5000~\cite{fischer2008nek5000} and (the recent) Neko~\cite{jansson2021neko}. While prior works have shown FPGAs to outperform CPUs (and, in some cases, GPUs) on a subset of larger applications (smaller kernels, e.g.~\cite{sano2013fpga,podobas2017designing,karp2021High}), we map (for the first time to our knowledge) the entire SEM-based solver to a modern FPGA, including analyzing and modeling its data movement properties from the perspective of reconfigurable systems.

We claim the following contributions:
\begin{itemize}
\item We leverage High-Level Synthesis (HLS) to create the first (to our knowledge) full SEM-based Pressure-Poisson solver on modern FPGAs targeting both single- and double-precision,
\item We propose an algorithm on FPGAs that trades more compute for less memory operations to alleviate the memory-boundness by computing geometric factors on the fly,
\item We develop a performance and cost model for generic I/O data movement for our solver,
\item We empirically evaluate the performance, the area, and the power-consuming properties of our accelerator design and position it against existing CPUs, and
\item We reveal new opportunities and challenges for using reconfigurable systems in CFD applications in the future.
\end{itemize}

\section{Field-Programmable Gate Arrays}

A Field-Programmable Gate Array~\cite{kuon2008fpga} is a type of programmable logic device whose logical functionality is (unlike, for example, a general-purpose CPU) flexible  after it has been manufactured. This is achieved by providing hundreds of thousands of reconfigurable look-up tables (LUTs), onto which logic can be dynamically mapped. The amount of LUTs that are occupied by a particular hardware design is called \textit{logic utilization}. Connectivity between LUTs is provided by a highly reconfigurable and versatile routing network. The flexible interconnect and LUTs are what made FPGAs an attractive vehicle for developing and testing new hardware. However, compared to ASIC designs, FPGAs often run at an order of magnitude lower frequency (e.g., often in the range 100-500 MHz contra 3-4 GHz of CPUs).
Early FPGAs only had LUTs and a flexible interconnect, but it was soon discovered that certain types of circuits mapped poorly to said structures and often consumed the entire FPGA (e.g., multiplications). This led to the inception of dedicated ASIC blocks called \textit{Digital Signal Processing (DSP)} blocks that handled these expensive operations. Today, DSP blocks are even capable of executing single-precision multiply-accumulate operations and are the reason FPGAs can compete with CPUs and GPUs in terms of computing performance and be more power-efficient (up to 22.3x more energy efficient compared to alternatives for e.g., image processing~\cite{qasaimeh2019comparing}). Finally, modern FPGAs include on-chip SRAM called \textit{Block RAM} (BRAM), providing a flexible high-bandwidth storage on-chip, which is typically on the order of hundreds of Mbit on high-end FPGAs.

Using FPGAs has historically been a very tedious and complex exercise that came with a steep learning curve. Complex EDA tools were often used, and designs were described in low-level hardware description languages, such as VHDL or Verilog. In the early 2010s, the increased maturity in High-Level Synthesis (HLS) facilitated a greater adoption and research of applying FPGAs for HPC. HLS tools enabled the use of abstract programming languages to describe hardware. Today, HLS tools exist for several programming languages and parallel models, such as C/C++~\cite{pilato2013bambu,canis2011legup}, OpenCL~\cite{czajkowski2012opencl}, OpenMP~\cite{podobas2016empowering}, CUDA~\cite{papakonstantinou2009fcuda}, and even Java~\cite{becker2015maxeler}. In this particular study, we use HLS as a method for creating a custom accelerator for the \textit{spectral element method}.

\section{The Spectral Element Method} 
The spectral element method has been widely acclaimed for its accuracy and scalability. In this section, we will briefly cover the algorithmic aspects of SEM and how we use it to discretize the Poisson equation in particular. Since the computation of the pressure in incompressible flow corresponds to the Poisson equation, and as it is the main source of stiffness, it frequently dominates the compute time of solvers such as Nek5000 and its successors NekRS \cite{fischer2021nekrs} and Neko. By solving the Poisson equation we, therefore, capture the core of the entire solver. The Poisson equation with homogeneous zero boundary conditions on a domain $\Omega$ can be written as 
\begin{equation}
\begin{split}
    \nabla^2 u &= f, \quad u \in \Omega,\\
    u &= 0, \quad u \in \partial\Omega.
\end{split}
\end{equation}
To discretize the system with SEM we introduce the weak form of the Poisson equation; find $u \in V \subset H^1_0$ s.t.

\begin{equation}
    \int_\Omega \nabla u \nabla v d \Omega = \int_\Omega f v d\Omega, \quad \forall v \in V.
\end{equation}
As SEM is a finite element method with high order basis functions we start our discretization by decomposing the domain $\Omega$ into $E$ non-overlapping elements $ \Omega = \bigcup_e^E \Omega^e$. We then approximate the solution by instead of using the continuous space $V$ we use a discretized space $V^N$ on a reference element with basis functions $l_i$. For basis functions, we use the $N$th order Legendre polynomials $L_N$ interpolated on the Gauss-Lobatto-Legendre (GLL) quadrature points $\xi_i$. The number of GLL points corresponds with the polynomial order as $N+1$. With these basis functions,  the local solution $u^e$ on a hexahedral reference element can then be expressed as a tensor product according to
\begin{equation}
    u^e(\xi, \eta, \gamma) = \sum_{i,j,k}^N u^e_{i,j,k} l_i(\xi)l_j(\eta)l_k(\gamma)
\end{equation}
where we have introduced the weights $u^e_{i,j,k}$ and where $\xi, \eta, \gamma$ correspond to the position in the reference element. Using this discretization we can then rewrite the weak form of the Poisson equation to the following bilinear form 
\begin{equation}\label{eq:sem_disc}
    \sum^E_e (v^e)^T \mathbf{D}^T \mathbf{G}^e\mathbf{D} u^e = \sum^E_e (v^e)^T A^e u^e
\end{equation}
where we introduce the tensor $\mathbf{G^e}$ which contains the geometric data mapping the actual element to the reference element and the differential matrix $\mathbf{D}$. Of importance for us is that forming this system is incredibly expensive\cite{deville2002high}. It is therefore evaluated in a matrix-free fashion.

\subsection{Matrix-Free Evaluation of $Ax$}
As the entire discrete matrix shown in \eqref{eq:sem_disc} is very expensive to form we instead split the system into a local computation of $A_{L}$ and communication along element boundaries. This communication is done in a \textit{gather-scatter} phase and our system can then instead be described as
\begin{equation}
    \sum^E_e (v^e)^T A^e u^e = (\mathbf{Q}v)^TA_{L}\mathbf{Q}u
\end{equation}
where the matrices $\mathbf{Q}, \mathbf{Q}^T$ are never explicitly formed, but only the operation they do is performed in the gather-scatter phase. With the scatter operator we also introduce the local and global representation of $u$, $u_L=\mathbf{Q}u$. Going forward we will always use the local representation $u_L$, replicating the values along element boundaries. To avoid cluttering we will drop the subscript $u_L$, as all vectors in the continued discussion will have replicated data along the element boundary. The matrix-free evaluation of the discrete system is described in depth in \cite{deville2002high} and is key to the high parallelism and scalability of SEM. Throughout this paper, when we mention $Ax$ in the context of spectral elements we then really talk about the operation $Ax = \mathbf{Q}\mathbf{Q}^TA_{L}x$ where $\mathbf{Q}\mathbf{Q}^T$ is the gather-scatter and $A_L$ is the local evaluation of the system.  In previous works only the local evaluation, $A_Lx$, has been considered on FPGAs.

\subsection{The Conjugate Gradient Method}
The preconditioned conjugate gradient (CG) method is one of the most common methods to solve positive-definite linear systems. In our work, we consider the unpreconditioned system as a first step towards using FPGAs for high fidelity computational fluid dynamics. We show the pseudocode for unpreconditioned CG combined with SEM in Algorithm \ref{alg:cg} where we denote arrays at iteration i with $x^{(i)}$ and scalars with greek letters and subscript i.e. $\rho_i$. One thing to note is that since we operate on the local arrays, and the boundary values are therefore duplicated, we need to introduce the vector $c$ to perform the correct reductions. Of importance for our further analysis is the prevalence of global reductions/\textit{synchronization points} on lines 8 and 11. These computations impose constraints on any implementation of CG as the computation cannot progress further until these scalars have been computed. Compared to other Krylov methods, one of the attractive features of CG is its relatively simple implementation and small memory footprint. For an overview of CG and many other iterative methods please see \cite{barrett1994templates}.
\begin{algorithm}[t]
  \begin{algorithmic}[1]
    \State{$r^{(0)} \gets b - Ax^{(0)}$}
    \State{$\hat{\beta}^{(0)} \gets 0$}
    \State{$p^{(0)} \gets \vec{0}$}
    \For{$i = 1,2, \ldots$}
    \State{$p^{(i)} \gets r^{(i-1)} + \hat{\beta}_{i-1}p^{(i-1)}$}
    \State{$w^{(i)} \gets QQ^TA_L p^{(i)}$}
    \State{Mask boundary conditions on $w^{(i)}$}
    \State{$\alpha_i \gets \rho_{i-1}/\langle p^{(i)}, w^{(i)},c\rangle$}\vspace{2pt}
    \State{$x^{(i)} \gets x^{(i-1)} + \alpha_{i}p^{(i)}$}\vspace{2pt}
    \State{$r^{(i)} \gets r^{(i-1)} - \alpha_{i}w^{(i)}$}\vspace{2pt}
    \State{$\rho_{i} \gets \langle r^{(i)}, r^{(i)},c\rangle$}
    \State{$\hat{\beta}_{i} \gets \rho_i/\rho_{i-1}$}
    \State{Check $\rho_i$ for convergence.}
    \EndFor
    
  \end{algorithmic}
  \caption{\label{alg:cg} Unpreconditioned CG applied on SEM.}
\end{algorithm}

\section{Theoretical Performance Analysis}
When designing a custom accelerator, such as the SEM solver accelerator we incept in this work, it is crucial to first derive a performance model for said computation. An analytical performance model can help designing an accelerator in multiple ways: \textbf{(i)} we can easier understand the bottlenecks of the application, \textbf{(ii)} we can derive the theoretically observable peak performance and also use it to model future (today non-existing) architecture (e.g.,~\cite{zohouri2018combined,karp2021High}, and \textbf{(iii)} designing an accelerator is a tough optimization problem where we want to optimize performance given certain area constraints, and a model can help driving decisions in the right direction.

Therefore, we will first develop and consider the theoretical performance of our CG solver and our SEM discretization, before going on to the implementation. We know beforehand that CG is notoriously memory-bound~\cite{marjanovic2014performance}, and  we aim to link our developed theory with the hardware parameters of the reconfigurable architecture. We base our model from previous findings by V. Elango et al. \cite{elango2014characterizing} for the optimal I/O~\footnote{We will adapt a fairly general notion of I/O, which include any traffic outside the processing core, in particular, we focus on accesses to DRAM/External memory (the \textit{vertical} I/O cost).} cost for CG (without preconditioning). We will extend their work to assess the I/O cost of the SEM discretization, which in turn will allow us to evaluate how memory-bound applications (in this case, SEM) can benefit from custom hardware such as FPGAs. Furthermore, we will also cover the expensive gather-scatter operation (including boundary condition masking) and its impact on performance.

\subsection{Machine Model}
Our machine model for this work will be that of a memory machine with unlimited slow memory and with a small fast memory of size $S$. This model corresponds well to an abstract general-purpose processor (CPU), graphics processing unit (GPU), or FPGAs, where the $S$ memory corresponds to caches (CPUs/GPUs) and Block RAM (FPGAs), and the slow memory is the external DDR or High-Bandwidth Memory (HBM). This abstract machine model can then be used to reason around the I/O cost of a program. This machine model has in particular been used in conjunction with the red-blue pebble game to derive various bounds on the I/O-cost as first introduced by Hong and Kung~\cite{jia1981complexity}. As CG is primarily bound by data movement, minimizing the I/O cost is at the core of our optimization process. In our work, we will use previous results obtained regarding the I/O cost of CG, to optimize and evaluate our FPGA implementation of CG and SEM discretization.

\subsection{I/O cost for the Conjugate Gradient and Spectral Element Method}
We will now go on to present our analysis of the I/O cost of SEM and unpreconditioned CG.  To begin our discussion we first note that an I/O lower bound for vertical data movement of the unpreconditioned CG was obtained in the excellent work by V. Elango et al. in \cite{elango2014characterizing}. In their work, they presented the I/O bound for unpreconditioned CG for many processors when the problem size is much larger than the small memory. For our intents and purposes we will also consider the small memory, but focus on only one processing element and rewrite their results as the following:

After iteration $i$ of unpreconditioned CG, the I/O cost $Q$ is lower bounded by
\begin{equation}
    Q \geq i(6n-4S)
\end{equation}
where $n$ is the vector length and $S$ is the size of the small fast memory. To understand this result further and its proof, please see section 5.2 in \cite{elango2014characterizing}. Of importance to our discussion is that in this particular case, V. Elango et al. considered recomputation of an array to be disallowed, and the computation of $Ax$ to be almost free with regards to data movement, meaning that the I/O cost of evaluating the discrete system $w=Ax$ is equal to the read from $x$ and the write to $w$. This is not true in the case of SEM as the computation of $Ax$ both depends on the geometric factors as well as the gather-scatter operation as we described previously. We, therefore, have several more loads and stores that we need to perform to evaluate $Ax$. As we need to load six geometric factors per point at every iteration we get that the lower bound for SEM instead would be
\begin{equation}\label{eq:sem_io_bound}
    Q_{\text{SEM}} \geq  i(6n+6n-4S) = i(12n-4S)
\end{equation}
where $n = E(N+1)^d$ is the number of points, $E$ is the number of elements, $N$ is the polynomial order, and $d$ is the dimension of the domain.  An interesting note here is that while the number of points $n = E(N+1)^d$ holds for any current implementation of SEM, in reality, the actual number of degrees of freedom are closer to $EN^d$. Because of the matrix-free approach, several points are duplicated along element boundaries. An insight from this bound is that a refactorization of the algorithm could thus potentially decrease the I/O cost further by not duplicating the points. This line of thought with considering the cost of $Ax$ and the actual number of degrees of freedom leads us to another observation that can be applied more generally. For unpreconditioned CG with arrays of length $n$, the I/O cost would be lower bounded by
\begin{equation}\label{eq:io_general_bound}
    Q_{\text{CG}} \geq i(6n+n_A-4S)
\end{equation}
where  $n_A$ is the minimum I/O cost needed to evaluate the the system $A$, given that we disallow recomputation. However, from an I/O perspective, if the discrete system $Ax$ incurs fewer than $n$ loads and stores it might be beneficial to evaluate $w^{(i)}=Ap^{(i)}$ on the fly rather than storing $w^{(i)}$ for further computations. This idea of trading computation for a lower I/O cost is also an interesting direction as we go forward with our implementation.

Now that we have covered the theoretical cost of our algorithm and different discretizations in general we must once again point out that it might be unfeasible to achieve an implementation close to our theoretical I/O cost (the issue of obtaining optimal schedules from lower bounds was recently discussed in \cite{Kwasniewski_2021}). While this is true we still think it is important to know what could be achieved. Without this knowledge, the evaluation of any implementation is largely based on intuition. We often showcase results in terms of how much better one is than the state-of-the-art or comparing to a roofline, but of maybe larger importance is how far away an implementation is from some theoretical optimal program. While we did not prove any lower bounds in this section we based our reasoning on previous results and we believe that this line of thought is necessary to achieve performant software on the exascale. As for the results presented in this section, we should also note that there is more to the evaluation of $Ax$ than the geometric factors, it also includes the gather-scatter operation $QQ^T$ which we have currently assigned an I/O cost of 0. The reason we have not covered it is that it does not obey the same rules as the other statements in our CG solver, because of the unstructured nature of our problem the gather-scatter operation is more similar to a graph problem, and obtaining a clear I/O bound is not trivial.

\subsection{Gather-Scatter and Masking}
The gather-scatter operation is responsible for summing the values across element boundaries, thus maintaining $C_0$ continuity across elements. Looking at the array $w^{(i)}$ it adds any values $w_i^{(i)} \in w^{(i)}$ s.t. the points that share the same position in space, $x(w_j^{(i)}) = x(w_k^{(i)})$. While the operation itself is simple, the issue comes down to that the indices connected are not necessarily aligned. Ideally, these indices could simply be summed as $Ax$ is computed in a streaming fashion. However, this is generally not possible because of the unstructured nature of SEM. Assuming that we cannot simply stream the values along element boundaries we must then load and store each element boundary value. For a discretization with polynomial order $N$, on our hexahedral elements, we have that the number of boundary values can be computed as
\begin{equation}
    n_{\text{gs}} = n\frac{(N+1)^3-(N-1)^3}{(N+1)^3}.
\end{equation}
Assuming that we need to load and store each of these values this implies an I/O cost of $Q_{\text{gs}} = 2n_{gs}$ for the gather scatter operation. Overall, for $N=7$, this leads to $Q_{\text{gs}} \approx n$. This impact is therefore not huge, but the problem is the unaligned accesses. As this is an unstructured problem we cannot guarantee any particular ordering or spatial locality of the elements in the solution vector, posing a potentially significant performance impact. The same argument can be applied for the boundary conditions, but as the number of values along the boundary grow with the surface of $\partial \Omega$ rather than the volume of $\Omega$ their impact for reasonable problem sizes is negligible ($n_{\text{bc}} \propto n^{2/3}$). We will therefore not consider it in our performance analysis, but rather assume all elements are connected on all sides in the gather-scatter operation.

\section{Accelerator Implementation}
In this section, we will present our accelerator implementation and design considerations. As our goal is to minimize the I/O cost we will continually relate the performance to our previous theoretical results and let them guide us in our design. Our implementation is made with Intel OpenCL SDK for FPGAs~\cite{czajkowski2012opencl}, a High-Level Synthesis Tool for Intel FPGAs. We start from our previous work where we implemented the local $A_Lx$ operation without the gather-scatter operation, in that work we also go into more detail on more practical optimization techniques \cite{karp2021High}. For general optimization guidelines for FPGAs, see for example \cite{ke-19a,de2020transformations}. In this work, we go on to extend our implementation of $A_L$ and implement the entire unpreconditioned CG solver on an FPGA, and focus on $N=7$ in particular. We choose $N=7$ as Nek5000 and its descendants are most commonly run with a polynomial orders between seven and eleven.

\subsection{Minimizing the I/O cost}
For the CG method when applied to SEM, as we show in Algorithm \ref{alg:cg}, one of the most important aspects to decrease the I/O cost is to not load any array between our synchronization points/reductions more than once. We accomplish this by loop fusion, fusing lines 5-6 and 9-11, lowering our I/O cost. However, as shown in Algorithm \ref{alg:cg} because of the matrix-free evaluation we have introduced a vector $c$ not to count values several times when making the reductions. In addition, we calculate $x^{(i)}$ at each iteration at a cost of $3n$ and execute the gather-scatter operation costing us $Q_{gs}$. As the gather-scatter and masking of boundary conditions operation need to finish before we start the reduction on line 7, this reduction can not be fused with the evaluation of $Ax$ and then also incurs an I/O cost as we reload $p^{(i)},w^{(i)}$. Overall, using the cost from \eqref{eq:sem_io_bound} with $S \gg n$ and taking $c$, the computation of $x$, as well as the new cost of the reduction at line 7 and gather-scatter into consideration, the I/O cost $Q_{\text{Ax-CG}}$ of our implementation is
\begin{equation}\label{eq:cg_cost}
    Q_{\text{Ax-CG}} = (12n + 3n + 3n + 2n + 2n_{gs})i = (20n + 2n_{gs})i.
\end{equation}
As for the number of computations per iteration, the vast number of flops can be attributed to the local $A_Lx$ computation totaling
\begin{equation}
    W_{A_Lx} = n(12(N+1)+15).
\end{equation}
For the entire solver, we then get (omitting the gather-scatter operation) that the total number of flops is
\begin{equation}\label{eq:org_ax_cost}
    W_{\text{CG}} = n(12(N+1) + 25).
\end{equation}
However, looking at the results from equation \eqref{eq:io_general_bound} we see that decreasing the I/O cost of $Ax$ can decrease the total I/O cost tremendously. As we precompute $\mathbf{G}$ to then compute $Ax$, we could therefore reduce the I/O cost of $A_L$ by recomputing $\mathbf{G}$ directly from the information stored in the mesh at each iteration instead. As the geometric factors are only dependent on the mesh, we could tremendously decrease the I/O cost as only eight values (compared to $6(N+1)^3$) would then be needed per element.  We will refer to this idea of recomputing values on the fly to obtain a lower I/O cost as \textit{rematerialization} as the principles behind it are similar to the compiler technique with the same name \cite{briggs1992rematerialization}. For the computation of $G$ though, we need to compute the Jacobian inverse, $J^{-1}$, and as this involves division the performance penalties are large. However, we can limit ourselves to only precompute $J^{-1}$ and computing $G$ on the fly otherwise. This means that we only need to load $J^{-1}$ instead of the six geometric factors leading to an I/O cost of $n$ instead of $6n$ for the computation of $Ax$ and a total cost of
\begin{equation}\label{eq:remat_cost}
    Q_{\text{CG-Remat.}} = (7n + 3n + 3n + 2n + 2n_{gs})i = (15n + 2n_{gs})i.
\end{equation}
This causes the number of computations to increase tremendously though.  We should also note that in our implementation we limit ourselves to meshes with hexahedral elements that are not curved but only linearly deformed. For more general meshes with curved elements, rematerialization would be even more expensive.  For the whole CG solver we now instead have (omitting computations that take place less than $n/(N+1)$ times)
\begin{equation}\label{eq:remat_flops}
    W_{\text{CG-Remat.}} \approx n(30(N+1) + 106).
\end{equation}
For $N=8$ we then have a more than 3-fold increase in the number of computations compared to \eqref{eq:org_ax_cost}. However, we should note that this is a rather naive approach and there may be algorithmic improvements that alleviate the increase in floating point operations partially. As the amount of computation necessary increases, we expect that rematerialization can be a potential alternative when the machine imbalance is large. As the operational intensity for $A_Lx$ increases to $I = (30(N+1) + 96)/3 \implies I = 112$ flop/word for $N=7$ compared to $I = (12(7+1)+15)/8=13.8$ flop/word for the original version. We can then clearly see that this operation is only relevant when the amount of computing power compared to the bandwidth is large. As recent Intel FPGAs such as the Intel Stratix 10 have tremendous computing power for single-precision computations we believe that rematerialization may be beneficial for this FPGA when running the solver with FP32.

\subsection{Maximizing Memory Bandwidth}\label{subsec:memband}
To utilize the external memory bandwidth as efficiently as possible, we manually place the different arrays on different DRAM memory banks. This enables us to saturate the memory ports to each bank as the local computations in $A_Lx$ as well as all the vector additions and reductions have a high degree of spatial locality, enabling large coalesced accesses to DRAM each cycle. There is an issue of balancing the arrays between the memory banks to enable complete utilization though. As for the operation $A_Lx$, we can easily balance the 8 arrays among the memory banks between the 4 banks, but as the reductions and vector operations in the solver cannot be as easily split between the different memory banks we cannot saturate the entire memory bandwidth. By examining the algorithm and carefully placing all the vectors on suitable memory banks we expect to achieve  $62\%$ of the theoretical bandwidth $\beta$ for the original formulation and $55\%$ for the rematerialization formulation.

As for the nonaligned nature of the gather-scatter operation, this poses a severe problem. The current code structure makes a nonaligned load and store for each boundary value and can thus only utilize a fraction of the global memory bandwidth. As we are mimicking the code structure of previous GPU and CPU implementations this means that we need to execute $2n_{gs}$ unaligned loads and stores. Related work~\cite{me-ke-20a} evaluating fully random access patterns with nonaligned loads and stores has shown a performance of around 60M transactions per second per DDR memory bank on the Stratix 10 architecture, corresponding to around 5 clock cycles per pair of read and write operations relative to the 300MHz of the memory interface. 

In the gather-scatter operation for SEM, the pattern is not fully random, but also not strictly pairwise, as multiple reads have to be completed before the sums are written back to the respective locations. This slight non-randomness means that roughly 1/3 of the loads and stores are aligned in memory and can be coalesced. For fully random accesses we observe that it takes around 3 cycles per read/write, but as one-third of the loads and stores are aligned, we can load eight values at the time one-third of the time leading to that it only takes $3\cdot(2/3\cdot1/(3\cdot8)) = 2.125$ cycles per load/store operation. However, this difference compared to aligned accesses still greatly impacts the runtime of the kernel and assuming we have a bandwidth of $\beta_{eff}$ words every clock cycle for the rest of the solver this means that our effective utilization of only one value per every $2.125$ cycles in the gather-scatter phase can have a large impact. We, therefore, introduce a model parameter $\beta_{gs}$ which is the number of words per cycle loaded in the gather-scatter phase. The modeled computation time can then be expressed as the following for a CG implementation with I/O cost $Q_{CG}$
\begin{equation}\label{eq:modeltime}
    T_c = \frac{Q_{CG}-2n_{gs}}{\beta_{eff}} +\frac{2n_{gs}}{\beta_{gs}}.
\end{equation}
This goes to show how impactful the gather-scatter is if we cannot make aligned loads and stores on the FPGA. As the bandwidth $\beta_{gs}$ is not affected by word length, this also implies that the performance benefits of moving to lower precision because of the larger $\beta$ might not be as significant as one might expect. 
\section{Experimental Setup}
For our measurements, we used the Noctua Cluster at Paderborn Center for High-Performance Computing. In particular, we used Bittware 520N cards equipped with an Intel Stratix 2800 GX FPGAs and 4 banks of DDR-4 memory clocked at 300 MHz and with a memory interface of 512 bits each, giving us a $\beta=32$ words/cycle for double precision and 64 for single precision. In GB/s the theoretical peak bandwidth $B$ is therefore 76.8 GB/s. For the CPU measurements we used the Beskow supercomputer at KTH, a Cray XC40 equipped with 2x 16 core Xeon E5-2698v3 Haswell CPUs clocked at 2.3 GHz per node. The DRAM bandwidth has previously been measured to a STREAM Triad bandwidth of 90 GB/s giving us a $\beta \approx 5$ words/cycle for double precision and 10 for single precision. We will use the clock frequency of 2.3GHz for our results, but it should be noted that there may be a slight variation because of Intel Turbo Boost technology. For the measurements, we parallelized the code with MPI over all 32 cores. For both the FPGA and CPU measurements, we made use of the pre-release of the spectral element framework Neko\cite{jansson2021neko}. For the FPGA we use Intel OpenCL SDK version 20.2 and Quartus Prime version 19.4 as well as the GNU compiler version 10.2.0. To interface with Fortran, we utilized CLFORTRAN developed by Company for Advanced Supercomputing Solutions\cite{CLFORTRANrepo}. We have also made our implementation available online\footnote{https://github.com/ExtremeFLOW/poisson\_fpga}. For the Haswell CPUs, we used the Intel Compiler version 19.1.1.217 and cray-mpich version 7.7.16. We also made performance and power measurements on a Marvell ThunderX2 (TX2) with 32 cores clocked at 2.2GHz at the PDC Center for High Performance Computing. For the TX2 we used GCC 9.2.0 and obtained a STREAM Triad bandwidth of $B=108$ GB/s. For the power measurements on the TX2, we used the Marvell tx2mon kernel module~\footnote{https://github.com/Marvell-SPBU/tx2mon}. For the FPGA power measurements, we used Bittware provided MMD functions that can be accessed through an API in OpenCL. We were unable to make any power measurements on the Cray XC40 nodes.

To assess the performance, we solved the Poisson equation on a cubic domain for different numbers of elements (128-32768). To preserve the generality of the code for different meshes, we did not make any assumptions on the geometry other than that we only had linear deformations for the rematerialization. 
\section{Results}
In this section, we will use our analytical bounds and modeled performance to assess our FPGA implementation of SEM. We will also consider a CPU baseline in Neko and use our performance model from \cite{jansson2021neko} to contrast our FPGA performance with a state-of-the-art CPU implementation. 
\subsection{$A_Lx$ with Rematerialization}
\begin{figure*}
    \centering
    \begin{subfigure}[t]{0.5\textwidth}
        \centering
        \includegraphics[scale=0.6]{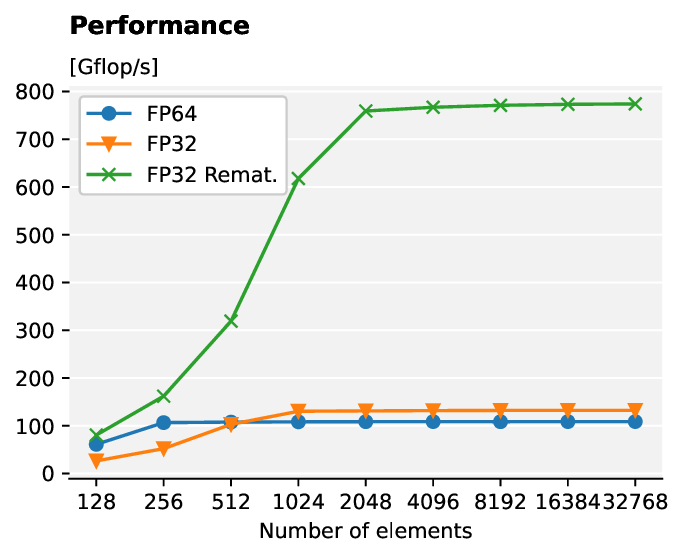}
    \end{subfigure}%
    ~ 
    \begin{subfigure}[t]{0.5\textwidth}
        \centering
        \includegraphics[scale=0.6]{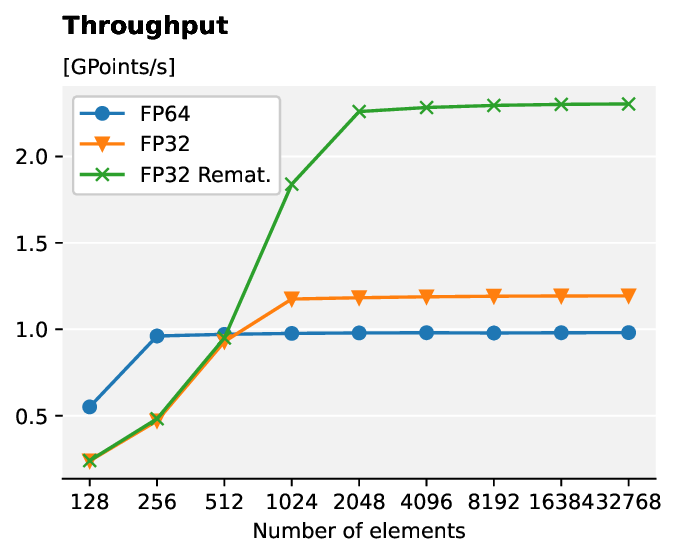}
    \end{subfigure}
    \caption{Performance for the $A_Lx$ kernel in FP64, FP32, and our new variation where we rematerialize the geometric factors. On the left, we show the raw performance in Gflop/s, and on the right, we show the number of points in the result vector, $w = A_Lx$, computed per second.}
    \label{fig:Ax_perf}
\end{figure*}

In our optimization process, we suggested computing the geometric factors on the fly if the amount of computing power is large. For our FPGA we projected that this could yield a performance improvement and lower runtime for single precision and we show the raw performance numbers in Figure \ref{fig:Ax_perf}. We greatly increase the performance compared to previous implementations and expect to reach even higher if we can increase the clock frequency of the kernel. With regards to total runtime, the computation with rematerialization performs around 2x faster than the original version and requires around 33\% fewer cycles to make the computation for inputs larger than 2048 elements. We expect that this optimzation can also be of benfit for other similar discretizations and for computing units with a high machine imbalance. The Haswell CPU does not have a large enough amount of compute for this to beneficial though. Of note for our FPGA is that we can saturate the available memory bandwidth for double-precision much earlier than for single-precision, something that was also noted in our previous work and shown in the appendix to \cite{karp2021High}.
\begin{table}[t]
{\centering
\begin{tabular}{l ll l l }\toprule
\multirow{2}{*}{\textbf{Version}} & $f_{\text{max}}$  & \textbf{Logic Util.} &  \textbf{BRAMs} & \textbf{DSPs} \\

 & (MHz) &  (\%)  & (\%) & (\%)  \\ \midrule
FP32-$A_Lx$-Remat. & 191   &  34   & 40 & \color{orange}{\textbf{71}}\\
FP32-$A_Lx$ & 150 & 29 & 43 & 34    \\
FP64-$A_Lx$ & 274  & 67 & 42  &  41  \\\hline
FP32-CG-Remat. &  156 & 32 & 55 & \color{orange}{\textbf{74}}\\
FP32-CG &  292 & 30 & 39 & 28    \\
FP64-CG &  204 & 59 & 52&35    \\\bottomrule
\end{tabular}}
\caption{Synthesis results for our designed accelerators in terms of operating frequency ($f_{max}$), logic utilization, on-chip storage (BRAM), and compute resources (DSPs). Note that the HLS tool and its backed was unable to increase the $f_{max}$ for some of the kernels and that further performance can be gained by manually reducing the critical path. Notice the increased DSP utilization for the rematerialized versions. }\label{tbl:resources}
\end{table}

\begin{table}[t]

{\centering
\begin{tabular}{l l ll l }\toprule
\textbf{Version} & $\beta_{eff}$ &  Model $\beta_{eff}$& $\beta_{gs}$ & Model $\beta_{gs}$ \\ \midrule
FP32-CG-Remat. &  21.6 & 35 & 0.474 & 0.47\\
FP32-CG &  22.4 & 40 & 0.54 & 0.47   \\
FP64-CG &  16 & 20 & 0.53 & 0.47    \\\bottomrule
\end{tabular}}
\caption{The observed contra modeled external bandwidth in words/cycle as experienced by our SEM-based accelerator on both the solver and the gather-scatter phase.}\label{tbl:model_split}
\end{table}
\subsection{Solver Performance}
\begin{figure}
    \centering
    \includegraphics[scale=0.7]{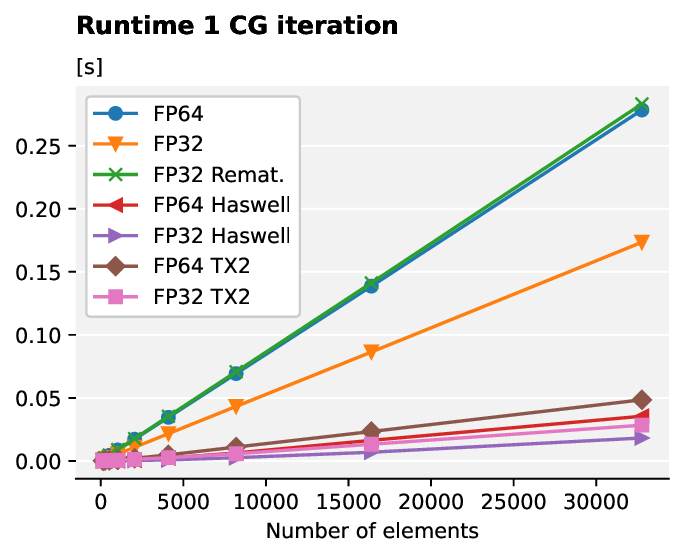}
    \caption{Total runtime of one CG iteration for the FP64, FP32, and the FP32 version with rematerialization on FPGA as well as FP64 and FP32 on the Haswell CPUs and Marvell ThunderX2.}
    \label{fig:cg_time}
\end{figure}

For the entire unpreconditioned CG solver with our SEM discretization we see in Fig. \ref{fig:cg_time} that the performance gain made from the new $A_Lx$ kernel with rematerialization of the geometric factors lead to potentially higher performance for single-precision, however, our implementation clocked in at a quite low frequency $156$MHz, hence the lower raw numbers compared to the achieved performance for only the $A_Lx$ kernel. Comparing the FPGA, Haswell, and the TX2 we see that the FPGA still has ways to go before competing with the two different CPUs, in particular because of the suboptimal gather-scatter. While the gather-scatter is the main difference, the memory system is also an important aspect. We should point out that we do compare one FPGA with one XC40 node, as the main memory system is a deciding factor of the performance. As the XC40 node's total memory bandwidth was measured to, ~90GB/s, it is also no surprise that it beats the FPGA. For the TX2 this is even more amplified as the STREAM bandwidth was measured to 108GB/s. However, it would appear the GCC compiler does not produce as efficient code for the TX2 as the intel compiler does for the Haswell. On the XC40 we get an exactly 2x performance increase when making single-precision computations compared to double-precision. On the TX2 the performance difference is much more modest, implying that it does not fully saturate the bandwidth. Another aspect we should mention is that our focus in this article is on large problem sizes, for CPUs the best performance is achieved when the potential data reuse in the cache is high, i.e. $S$ is comparably large compared to $n$. In the domain when $n\gg S$ we have that the FPGA performs 5$\times$ worse than the Haswell node, but more than $6 \times$ better than one Haswell core. We expect that the performance for large problem sizes will be even higher for e.g. GPUs with HBM2 than both our CPU and FPGA platforms. Looking at smaller problem sizes per FPGA may be an interesting domain to explore in the future as the FPGA enables us to completely control the usage of on-chip BRAM. 

\begin{figure*}
    \centering
    \includegraphics[scale=0.6]{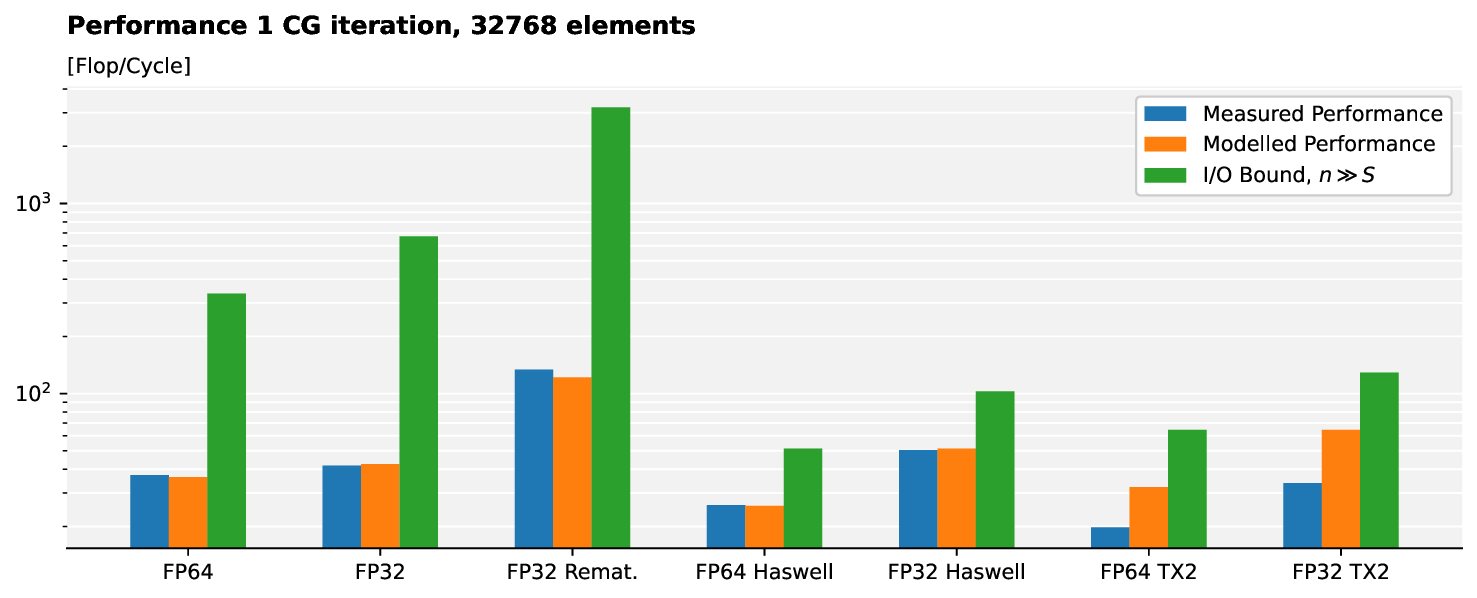}
    \caption{Performance in Flop/Cycle for the FP64, FP32, and the FP32 version with rematerialization as well as for FP64 and FP32 on one XC40 node and one Marvell ThunderX2. We obtain the modeled performance with \eqref{eq:modeltime} for the FPGAs and we use the Neko performance model for the CPUs based on their STREAM bandwidth. For the I/O bound performance we base this on the bandwidth $\beta$ of a given architecture and the I/O cost presented in equation \eqref{eq:io_general_bound}.}
    \label{fig:cg_perf}
\end{figure*}

Normalizing the CPU and FPGAs with regard to frequency we see that the FPGA is comparable or better than the CPUs per cycle (Fig.~\ref{fig:cg_perf}) and that the rematerialization performs better than the original single precision implementation. While it is unrealistic that the CPU and FPGA will run at the same frequency, normalizing the performance with regards to frequency makes it easier to compare FPGA versions with varying frequency. In the plot, we show the measured performance, as well as the modeled performance for each architecture and implementation by combining equations \eqref{eq:cg_cost} and \eqref{eq:modeltime}. In addition, we show the theoretical best performance according to our analysis, where we combine the cost from equation \eqref{eq:sem_io_bound} with the maximal bandwidth $\beta$ for each architecture. For the rematerialized version we use equation \eqref{eq:remat_cost} for the I/O cost and assess the theoretical limit with equation \eqref{eq:io_general_bound} and $n_A=n$. However, as can be seen, is that because of its higher frequency and cache, the CPU implementations are not at all as impacted by the fact that the gather-scatter kernel is limited to very few loads and stores per cycle. The relative impact compared to $\beta$ is not as large as it is for the FPGA. In addition, the cache exploits potential spatial locality that can be used. As for the modeled performance, we are within $10\%$ of the measured performance for all versions on the FPGA and within $2\%$ for the Haswell CPU. For the whole solver we achieved an effective bandwidth of $90-100\%$ of STREAM on Haswell, $60\%$ on the TX2, but significantly lower as can be seen in Table \ref{tbl:model_split} on the FPGA. 

Overall though, the FPGA performance model as shown in Fig. \ref{fig:cg_perf} is likely a bit too good to be true. Looking at Table \ref{tbl:model_split} we see that the performance model with our expected values of $\beta_{eff}$ and $\beta_{gs}$ overshoot and undershoot their values respectively. In particular, our utilization of the global memory bandwidth is worse than we suggested in subsection \ref{subsec:memband}. The gather-scatter operation seems to perform slightly better than expected though, compensating for the rest of the solver. All-in-all the measured results still confirm our analysis that the gather-scatter operation is incredibly expensive on FPGAs in its current form. If we could utilize the DRAM bandwidth more efficiently for the gather-scatter and decrease the I/O cost of the SEM-CG closer to the limit presented in \eqref{eq:io_general_bound}, the performance could increase tremendously.

\subsection{Resource Utilization and Power Consumption}
The last thing to note is the resource utilization for the different designs. Inspecting Table \ref{tbl:resources} we see how the resource consumption varies depending on the use of double or single-precision arithmetic. As the Intel Stratix 10 FPGA has DSP blocks tailored to single-precision these results show how this translates to a lot more efficient hardware. Another aspect we must also consider is the relation between a single kernel on the FPGA and synthesizing the whole Poisson solver. For the $A_Lx$ we can obtain very high performance and use a lot of resources, but to route it and use it in the entire CG solver, there are challenges as we need to decrease the resources necessary for the $A_Lx$ kernel to obtain a functioning design. One approach would be to use multiple FPGAs for different parts of the solver but as off-chip bandwidth is slower than on-chip this would in reality cripple the performance. Another aspect of these designs is the highly varying frequency. It is possible to increase it, but it is a time-consuming process. Another aspect we saw was that more complex designs could also sometimes malfunction for completely unrelated reasons. The CG version with rematerialization for multiple designs therefore often performed incorrect reductions, even if we only changed the evaluation of $A_Lx$. This also points to that while HLS makes FPGA programming easier, the reliability and design considerations necessary when using FPGAs are still large. While plenty of work goes on to show good performance of FPGAs for specific kernels for certain applications we in this work showcase that the step from a single high-performing important kernel to a whole application is considerable.

On the bright side, inspecting Table \ref{tbl:power}, we measure a mere \textbf{70-80} Watt of power consumption for our FPGA accelerator, which is considerably less than current CPUs and GPUs. However, it should be noted that even though the power consumption was lower, still the Flop/J is not yet on par with the TX2 we compare against. If the power consumption remains constant though as the gather-scatter operation is improved we expect our FPGA to outperform CPUs in this regard.

\begin{table*}[t]

{\centering
\begin{tabular}{l l l l l l }\toprule
 & FP64 FPGA &  FP32 FPGA& FP32 FPGA Remat. & FP64 TX2 & FP32 TX2 \\ \midrule
Power [W] &  78.7 & 75.6 & 76.8 & 133.5 & 138.1\\
Energy [kJ] &  21.8 & 13 & 21.7 & 6.7 & 4\\\bottomrule
\end{tabular}}
\caption{The power consumption for our FPGA accelerator compared to the ThunderX2 in terms of power consumption (Watt) and total energy consumption (in Joules) for the CG solver running 1000 iteration with 32768 elements. }\label{tbl:power}
\end{table*}
\section{Opportunities and Challenges}
In this section, we will discuss the different hurdles and opportunities we have uncovered as we made this first initial implementation of SEM on FPGAs. 

\subsection{Memory Bandwidth}
As can be seen, is that even as we minimize the I/O cost the other factor that goes into the runtime is the available memory bandwidth $\beta$. Even if the gather-scatter phase would not be our performance bottleneck, the low memory bandwidth of current FPGAs poses a challenge, and HBM2, recently available for Intel and Xilinx FPGAs, is key for FPGAs to compete for this type of application when the problem size is large. Another complementary challenge is that the memory controller does not make good use of interleaved memory and we need to place it manually on the memory banks\cite{8945518}. With manual interleaving or partitioning to 32 banks of HBM2-enabled FPGAs, the gather-scatter phase might become up to 32x faster than the current version with a single bank of DDR memory. In addition, our measurements show how the low frequency of the FPGA makes the impact of non-aligned reads and writes even more detrimental than for CPUs with a higher frequency and larger cache.

\textbf{Future opportunities:} The emerging availability of high-bandwidth memory on FPGAs could change the game for FPGAs, but the main challenge is how to distribute data across the many channels on these devices.

\subsection{On and Off-chip Memory control}
An opportunity that FPGAs offer is the unmatched control of the memory layout, both in DRAM or HBM, but also on-chip memory. Currently, our FPGA implementation runs similar to a GPU with one PE per FPGA, however, one may consider novel approaches where we parallelize over each memory bank instead, treating one connection to the global memory as one PE. One may argue that the single largest benefit of FPGAs is the free control over on-chip BRAM. It is therefore possible that decreasing the problem size so that all arrays can fit into BRAM and use multiple FPGAs as a way forward for higher performance. BRAM is also the key to achieving higher performance for the gather scatter kernel. As it in essence is a graph computation many of the methods applied to graph processing on FPGAs can be applied on the gather-scatter operation\cite{besta2019graph}. In particular, by pre-partitioning the domain into suitable chunks that fit into BRAM we expect that the gather-scatter performance can increase tremendously. For this, we expect that partitioning the elements with parMETIS or similar \cite{karypis1997parmetis} can be the key to make the decomposition performant. Alternatively, specific sorting strategies can be employed to increase and exploit the locality in the gather-scatter phase, as discussed by Barrio et al.~\cite{ba-ca-14a}. With this type of improvement in place, we expect that the gather-scatter bottleneck could be alleviated.

\textbf{Future opportunity:} By re-partitioning the gather-scatter operation into domain chunks that fit into BRAM, which will be more amendable in emerging 3D stacked high-storage chips, we hypothesize that future emerging reconfigurable systems will overcome the current gather-scatter bottleneck that we found through this paper.

\subsection{Synchronization Points}
A challenge that arises because of the structure of the CG method is the synchronization points in the form of global reductions that we mentioned earlier. One aspect of FPGAs often praised is the unique opportunity to utilize available silicon efficiently. Synchronization points by their very nature pose a large issue for this notion as they essentially form a barrier that blocks us from using the entire chip at once. For example, in our case, the resource allocated to $A_Lx$ are unused more than half of the time because of the synchronization points. As GPUs and CPUs essentially implement very many small kernels an open question is whether FPGAs are suited for compute-heavy tasks with synchronization points. While our work implies that this might be a fundamental issue currently, we can not rule out that with the advent of 3D stacked BRAM and/or faster configuration times that FPGAs will not form serious contenders to other architectures even when synchronization points and large input sizes are considered.

\textbf{Future opportunity:} Synchronization points effectively hinder us from using the whole FPGA at once and multi-contexting on FPGAs is today very expensive, as reconfiguring an FPGA takes orders of seconds. With future 3D stacking~\cite{cevrero2009using}, we believe that multiple contexts can be held on-chip, and new contexts can be loaded on the order of microseconds (or less), which could open up new opportunities for remedying the issue of global reductions and synchronization points that are essential in many modern iterative solvers.

\subsection{HLS and FPGA workflow}
What is clear from a developer perspective is that while FPGAs offer many interesting features, the workflow to use them is still quite constrained and poses may be the largest challenge for FPGAs to become mainstream. Currently, while HLS is a large improvement compared to Hardware Description Languages, even with an understanding of the hardware it is a long path to a high performing FPGA design. This may be expected, but even then the unpredictability of an FPGA design workflow with regard to resource utilization and frequency makes the design process hard and time-consuming. The implementation in this paper took the better part of June and July to implement and optimize, even with previous functioning designs for the $A_Lx$ kernel.

\textbf{Future opportunity:} HLS is often a time-consuming effort, and results are often not available to post place\&route, which can take hours. Better early models and feedback to users will be crucial in the future, or perhaps the use of more coarse-grained elements such as FPGA overlays~\cite{brant2012zuma} or templates~\cite{podobas2020template} for faster development cycles.

\subsection{I/O Lower bounds}
While we are still far from the theoretically achievable performance, we should make clear that achieving an "optimal" solver is our goal. As we have seen in this work, there are challenges, in particular with regards to the gather-scatter kernel, but even on CPUs, we have ways to go as shown in Fig. \ref{fig:cg_perf}. We believe that the consideration of I/O cost should be one of the main guiding principles in the application optimization process and in this work we show a path where we start from a theoretical analysis and then use this to understand and optimize our application. In this work, we primarily focused on vertical data movement, but a similar approach also needs to be made when considering multiple processes and nodes and communication between them, so-called horizontal data movement to both design algorithms and future hardware.

\textbf{Future challenge:} As machine imbalance grows, and memory-hierarchies grow deeper and more complex, it will be inevitable and crucial to consider I/O bounds to reason around and obtain higher performance. 

\subsection{Floating Point Precision}
On the last note, we must also mention the performance impact of floating-point representation. It is clear that regardless if we are memory or compute-bound, decreasing the floating-point precision can increase the performance. In the case of our Haswell CPUs, we see almost exactly 2x by simply going from double to single precision. While there is  work on obtaining high accuracy and performance with lower floating-point precision in various Krylov solvers, the users of scientific software are often cautious. Further studies into the actual physical implications of lowering the floating-point precision are therefore necessary. We see that FPGAs offer an unmatched testbed and opportunity for experimenting with new precision formats.

\textbf{Future opportunities:} The reduction in numerical representation can yield high-performance, more compute, and better bandwidth utilization (more values per unit BW), and using FPGAs gives the opportunity to empirically evaluate the use of lower precision for future CFD codes.
 
\section{Related Work}
Analysis of the I/O lower bound for unpreconditioned CG and various other I/O bounds were done in a series of works by V. Elango et al. \cite{elango2014characterizing,10.1145/2693656} and also combined with a roofline analysis. However, to our knowledge, we are the first to use these results to reason around the spectral element method. 

Recent work on sparse matrix-vector multiplications as a building block for the conjugate gradient method has shown that an FPGA attached to one DDR4 can achieve excellent performance up to the memory roofline here~\cite{ja-om-20a}. However, the presented design relies on a custom component that could not have been efficiently implemented using HLS-based design methods yet. In the work of Grigoras et al., sparse matrix multiplication for the finite element method was also accelerated on FPGAs and combined with a performance model to make projections for the Nektar++ solver\cite{7577352}. 

As for the implementation of the matrix-free evaluation of SEM on FPGAs, the core computation $A_Lx$ has been implemented on both Xilinx and Intel FPGAs \cite{karp2021High,9307119}. However, as these works only optimize the most compute-heavy kernel, and the Xilinx work only evaluates a very high polynomial degree, many of the issues brought up in this work were largely untouched. 

Another approach was implemented by Blanchard et al. \cite{blanchard2020fpga} where they focused on the mini app CMT-bone-BE, which is a compressible flow solver, but where they optimized a similar kernel to $A_Lx$. There, they offloaded the optimized kernel to the FPGA, achieving performance of 1.5x over \textit{one CPU core} for the whole application. Their results indicate that off-chip memory bandwidth (over PCI-E) quickly becomes a limiting factor.

As for an implementation using FPGAs for unstructured meshes, the closest related work has been done on the discontinuous Galerkin method for electrodynamics~\cite{8457652} and more recently on shallow-water simulations~\cite{ke-sh-21i}. A flow solver based on the finite volume method was also implemented in \cite{nagy2012fpga}.
\section{Conclusion}
In this paper, we have presented a thorough performance analysis of our unpreconditioned SEM solver on FPGAs. We show initial performance results and point towards a set of opportunities and challenges that exist for FPGAs for this type of application, in particular how data movement, global reductions, and floating-point precision can have a major impact on the utilization and performance of FPGAs. In future work, we also want to evaluate the impact a preconditioner would have on this application on FPGAs. We hope that our insights can be of help as we approach an even more heterogeneous exascale landscape.

\begin{acks}
Financial support was provided by the SeRC Exascale Simulation Software Initiative (SESSI) and by the DEEP-SEA project. The DEEP-SEA project has received funding from the European Union's Horizon 2020/EuroHPC research and innovation programme under grant agreement No 955606. National contributions from the involved state members match the EuroHPC funding. This work is partially funded by the Federal Ministry of Education and Research (BMBF) and the state of North Rhine-Westphalia as part of the NHR Program. The authors gratefully acknowledge the funding of this project by computing time provided by the Paderborn Center for Parallel Computing (PC\textsuperscript{2}). We performed experiments on resources provided by the Swedish National Infrastructure for Computing (SNIC), partially funded by the Swedish Research Council through grant agreement no. 2018-05973, at PDC Center for High Performance Computing.
\end{acks}

\bibliographystyle{ACM-Reference-Format}
\bibliography{main}


\begin{thebibliography}{51}


\ifx \showCODEN    \undefined \def \showCODEN     #1{\unskip}     \fi
\ifx \showDOI      \undefined \def \showDOI       #1{#1}\fi
\ifx \showISBNx    \undefined \def \showISBNx     #1{\unskip}     \fi
\ifx \showISBNxiii \undefined \def \showISBNxiii  #1{\unskip}     \fi
\ifx \showISSN     \undefined \def \showISSN      #1{\unskip}     \fi
\ifx \showLCCN     \undefined \def \showLCCN      #1{\unskip}     \fi
\ifx \shownote     \undefined \def \shownote      #1{#1}          \fi
\ifx \showarticletitle \undefined \def \showarticletitle #1{#1}   \fi
\ifx \showURL      \undefined \def \showURL       {\relax}        \fi
\providecommand\bibfield[2]{#2}
\providecommand\bibinfo[2]{#2}
\providecommand\natexlab[1]{#1}
\providecommand\showeprint[2][]{arXiv:#2}

\bibitem[\protect\citeauthoryear{Barrett, Berry, Chan, Demmel, Donato,
  Dongarra, Eijkhout, Pozo, Romine, and Van~der Vorst}{Barrett
  et~al\mbox{.}}{1994}]%
        {barrett1994templates}
\bibfield{author}{\bibinfo{person}{Richard Barrett}, \bibinfo{person}{Michael
  Berry}, \bibinfo{person}{Tony~F Chan}, \bibinfo{person}{James Demmel},
  \bibinfo{person}{June Donato}, \bibinfo{person}{Jack Dongarra},
  \bibinfo{person}{Victor Eijkhout}, \bibinfo{person}{Roldan Pozo},
  \bibinfo{person}{Charles Romine}, {and} \bibinfo{person}{Henk Van~der
  Vorst}.} \bibinfo{year}{1994}\natexlab{}.
\newblock \bibinfo{booktitle}{\emph{Templates for the solution of linear
  systems: building blocks for iterative methods}}.
\newblock \bibinfo{publisher}{SIAM}.
\newblock


\bibitem[\protect\citeauthoryear{Barrio, Carreras, López, Óscar Robles,
  Jevtic, and Sierra}{Barrio et~al\mbox{.}}{2014}]%
        {ba-ca-14a}
\bibfield{author}{\bibinfo{person}{Pablo Barrio}, \bibinfo{person}{Carlos
  Carreras}, \bibinfo{person}{Juan~A. López}, \bibinfo{person}{Óscar Robles},
  \bibinfo{person}{Ruzica Jevtic}, {and} \bibinfo{person}{Roberto Sierra}.}
  \bibinfo{year}{2014}\natexlab{}.
\newblock \showarticletitle{Memory optimization in {FPGA}-accelerated
  scientific codes based on unstructured meshes}.
\newblock \bibinfo{journal}{\emph{Journal of Systems Architecture (JSA)}}
  \bibinfo{volume}{60}, \bibinfo{number}{7} (\bibinfo{year}{2014}),
  \bibinfo{pages}{579 -- 591}.
\newblock
\showISSN{1383-7621}
\urldef\tempurl%
\url{https://doi.org/10.1016/j.sysarc.2014.07.001}
\showDOI{\tempurl}


\bibitem[\protect\citeauthoryear{Becker, Mencer, Weston, and Gaydadjiev}{Becker
  et~al\mbox{.}}{2015}]%
        {becker2015maxeler}
\bibfield{author}{\bibinfo{person}{Tobias Becker}, \bibinfo{person}{Oskar
  Mencer}, \bibinfo{person}{Stephen Weston}, {and} \bibinfo{person}{Georgi
  Gaydadjiev}.} \bibinfo{year}{2015}\natexlab{}.
\newblock \showarticletitle{Maxeler data-flow in computational finance}.
\newblock In \bibinfo{booktitle}{\emph{FPGA Based Accelerators for Financial
  Applications}}. \bibinfo{publisher}{Springer}, \bibinfo{pages}{243--266}.
\newblock


\bibitem[\protect\citeauthoryear{Besta, Stanojevic, Licht, Ben-Nun, and
  Hoefler}{Besta et~al\mbox{.}}{2019}]%
        {besta2019graph}
\bibfield{author}{\bibinfo{person}{Maciej Besta}, \bibinfo{person}{Dimitri
  Stanojevic}, \bibinfo{person}{Johannes De~Fine Licht}, \bibinfo{person}{Tal
  Ben-Nun}, {and} \bibinfo{person}{Torsten Hoefler}.}
  \bibinfo{year}{2019}\natexlab{}.
\newblock \showarticletitle{Graph processing on fpgas: Taxonomy, survey,
  challenges}.
\newblock \bibinfo{journal}{\emph{arXiv preprint arXiv:1903.06697}}
  (\bibinfo{year}{2019}).
\newblock


\bibitem[\protect\citeauthoryear{Blanchard, Stitt, and Lam}{Blanchard
  et~al\mbox{.}}{2020}]%
        {blanchard2020fpga}
\bibfield{author}{\bibinfo{person}{Ryan Blanchard}, \bibinfo{person}{Greg
  Stitt}, {and} \bibinfo{person}{Herman Lam}.} \bibinfo{year}{2020}\natexlab{}.
\newblock \showarticletitle{FPGA Acceleration of Fluid-Flow Kernels}. In
  \bibinfo{booktitle}{\emph{2020 IEEE/ACM International Workshop on
  Heterogeneous High-performance Reconfigurable Computing (H2RC)}}. IEEE,
  \bibinfo{pages}{29--37}.
\newblock


\bibitem[\protect\citeauthoryear{Brant and Lemieux}{Brant and Lemieux}{2012}]%
        {brant2012zuma}
\bibfield{author}{\bibinfo{person}{Alexander Brant} {and}
  \bibinfo{person}{Guy~GF Lemieux}.} \bibinfo{year}{2012}\natexlab{}.
\newblock \showarticletitle{ZUMA: An open FPGA overlay architecture}. In
  \bibinfo{booktitle}{\emph{2012 IEEE 20th international symposium on
  field-programmable custom computing machines}}. IEEE,
  \bibinfo{pages}{93--96}.
\newblock


\bibitem[\protect\citeauthoryear{Briggs, Cooper, and Torczon}{Briggs
  et~al\mbox{.}}{1992}]%
        {briggs1992rematerialization}
\bibfield{author}{\bibinfo{person}{Preston Briggs}, \bibinfo{person}{Keith~D
  Cooper}, {and} \bibinfo{person}{Linda Torczon}.}
  \bibinfo{year}{1992}\natexlab{}.
\newblock \showarticletitle{Rematerialization}. In
  \bibinfo{booktitle}{\emph{Proceedings of the ACM SIGPLAN 1992 conference on
  Programming language design and implementation}}. \bibinfo{pages}{311--321}.
\newblock


\bibitem[\protect\citeauthoryear{Brown}{Brown}{2020}]%
        {9307119}
\bibfield{author}{\bibinfo{person}{Nick Brown}.}
  \bibinfo{year}{2020}\natexlab{}.
\newblock \showarticletitle{Exploring the acceleration of Nekbone on
  reconfigurable architectures}. In \bibinfo{booktitle}{\emph{2020 IEEE/ACM
  International Workshop on Heterogeneous High-performance Reconfigurable
  Computing (H2RC)}}. \bibinfo{pages}{19--28}.
\newblock
\urldef\tempurl%
\url{https://doi.org/10.1109/H2RC51942.2020.00008}
\showDOI{\tempurl}


\bibitem[\protect\citeauthoryear{Butrashvily}{Butrashvily}{[n.d.]}]%
        {CLFORTRANrepo}
\bibfield{author}{\bibinfo{person}{Mordechai Butrashvily}.}
  \bibinfo{year}{[n.d.]}\natexlab{}.
\newblock \bibinfo{title}{{CLFORTRAN}}.
\newblock
\newblock
\urldef\tempurl%
\url{https://github.com/cass-support/clfortran}
\showURL{%
\tempurl}
\newblock
\shownote{Accessed: Aug. 27, 2021.}


\bibitem[\protect\citeauthoryear{Canis, Choi, Aldham, Zhang, Kammoona,
  Anderson, Brown, and Czajkowski}{Canis et~al\mbox{.}}{2011}]%
        {canis2011legup}
\bibfield{author}{\bibinfo{person}{Andrew Canis}, \bibinfo{person}{Jongsok
  Choi}, \bibinfo{person}{Mark Aldham}, \bibinfo{person}{Victor Zhang},
  \bibinfo{person}{Ahmed Kammoona}, \bibinfo{person}{Jason~H Anderson},
  \bibinfo{person}{Stephen Brown}, {and} \bibinfo{person}{Tomasz Czajkowski}.}
  \bibinfo{year}{2011}\natexlab{}.
\newblock \showarticletitle{LegUp: high-level synthesis for FPGA-based
  processor/accelerator systems}. In \bibinfo{booktitle}{\emph{Proceedings of
  the 19th ACM/SIGDA international symposium on Field programmable gate
  arrays}}. \bibinfo{pages}{33--36}.
\newblock


\bibitem[\protect\citeauthoryear{Cevrero, Athanasopoulos, Parandeh-Afshar,
  Skerlj, Brisk, Leblebici, and Ienne}{Cevrero et~al\mbox{.}}{2009}]%
        {cevrero2009using}
\bibfield{author}{\bibinfo{person}{Alesandro Cevrero},
  \bibinfo{person}{Panagiotis Athanasopoulos}, \bibinfo{person}{Hadi
  Parandeh-Afshar}, \bibinfo{person}{Maurizio Skerlj}, \bibinfo{person}{Philip
  Brisk}, \bibinfo{person}{Yusuf Leblebici}, {and} \bibinfo{person}{Paolo
  Ienne}.} \bibinfo{year}{2009}\natexlab{}.
\newblock \showarticletitle{Using 3D integration technology to realize
  multi-context FPGAs}. In \bibinfo{booktitle}{\emph{2009 International
  Conference on Field Programmable Logic and Applications}}. IEEE,
  \bibinfo{pages}{507--510}.
\newblock


\bibitem[\protect\citeauthoryear{Czajkowski, Aydonat, Denisenko, Freeman,
  Kinsner, Neto, Wong, Yiannacouras, and Singh}{Czajkowski
  et~al\mbox{.}}{2012}]%
        {czajkowski2012opencl}
\bibfield{author}{\bibinfo{person}{Tomasz~S Czajkowski}, \bibinfo{person}{Utku
  Aydonat}, \bibinfo{person}{Dmitry Denisenko}, \bibinfo{person}{John Freeman},
  \bibinfo{person}{Michael Kinsner}, \bibinfo{person}{David Neto},
  \bibinfo{person}{Jason Wong}, \bibinfo{person}{Peter Yiannacouras}, {and}
  \bibinfo{person}{Deshanand~P Singh}.} \bibinfo{year}{2012}\natexlab{}.
\newblock \showarticletitle{From OpenCL to high-performance hardware on FPGAs}.
  In \bibinfo{booktitle}{\emph{22nd international conference on field
  programmable logic and applications (FPL)}}. IEEE, \bibinfo{pages}{531--534}.
\newblock


\bibitem[\protect\citeauthoryear{de~Fine~Licht, Besta, Meierhans, and
  Hoefler}{de~Fine~Licht et~al\mbox{.}}{2020}]%
        {de2020transformations}
\bibfield{author}{\bibinfo{person}{Johannes de Fine~Licht},
  \bibinfo{person}{Maciej Besta}, \bibinfo{person}{Simon Meierhans}, {and}
  \bibinfo{person}{Torsten Hoefler}.} \bibinfo{year}{2020}\natexlab{}.
\newblock \showarticletitle{Transformations of high-level synthesis codes for
  high-performance computing}.
\newblock \bibinfo{journal}{\emph{IEEE Transactions on Parallel and Distributed
  Systems}} \bibinfo{volume}{32}, \bibinfo{number}{5} (\bibinfo{year}{2020}),
  \bibinfo{pages}{1014--1029}.
\newblock


\bibitem[\protect\citeauthoryear{Deville, Fischer, Fischer, Mund,
  et~al\mbox{.}}{Deville et~al\mbox{.}}{2002}]%
        {deville2002high}
\bibfield{author}{\bibinfo{person}{Michel~O Deville}, \bibinfo{person}{Paul~F
  Fischer}, \bibinfo{person}{Paul~F Fischer}, \bibinfo{person}{EH Mund},
  {et~al\mbox{.}}} \bibinfo{year}{2002}\natexlab{}.
\newblock \bibinfo{booktitle}{\emph{High-order methods for incompressible fluid
  flow}}. Vol.~\bibinfo{volume}{9}.
\newblock \bibinfo{publisher}{Cambridge university press}.
\newblock


\bibitem[\protect\citeauthoryear{Elango, Rastello, Pouchet, Ramanujam, and
  Sadayappan}{Elango et~al\mbox{.}}{2014}]%
        {elango2014characterizing}
\bibfield{author}{\bibinfo{person}{Venmugil Elango}, \bibinfo{person}{Fabrice
  Rastello}, \bibinfo{person}{Louis-No{\"e}l Pouchet},
  \bibinfo{person}{Jagannathan Ramanujam}, {and} \bibinfo{person}{Ponnuswamy
  Sadayappan}.} \bibinfo{year}{2014}\natexlab{}.
\newblock \showarticletitle{On characterizing the data movement complexity of
  computational DAGs for parallel execution}. In
  \bibinfo{booktitle}{\emph{Proceedings of the 26th ACM Symposium on
  Parallelism in Algorithms and Architectures}}. \bibinfo{pages}{296--306}.
\newblock


\bibitem[\protect\citeauthoryear{Elango, Sedaghati, Rastello, Pouchet,
  Ramanujam, Teodorescu, and Sadayappan}{Elango et~al\mbox{.}}{2015}]%
        {10.1145/2693656}
\bibfield{author}{\bibinfo{person}{Venmugil Elango}, \bibinfo{person}{Naser
  Sedaghati}, \bibinfo{person}{Fabrice Rastello},
  \bibinfo{person}{Louis-No\"{e}l Pouchet}, \bibinfo{person}{J. Ramanujam},
  \bibinfo{person}{Radu Teodorescu}, {and} \bibinfo{person}{P. Sadayappan}.}
  \bibinfo{year}{2015}\natexlab{}.
\newblock \showarticletitle{On Using the Roofline Model with Lower Bounds on
  Data Movement}.
\newblock \bibinfo{journal}{\emph{ACM Trans. Archit. Code Optim.}}
  \bibinfo{volume}{11}, \bibinfo{number}{4}, Article \bibinfo{articleno}{67}
  (\bibinfo{date}{Jan.} \bibinfo{year}{2015}), \bibinfo{numpages}{23}~pages.
\newblock
\showISSN{1544-3566}
\urldef\tempurl%
\url{https://doi.org/10.1145/2693656}
\showDOI{\tempurl}


\bibitem[\protect\citeauthoryear{Fischer, Kerkemeier, Min, Lan, Phillips,
  Rathnayake, Merzari, Tomboulides, Karakus, Chalmers, and Warburton}{Fischer
  et~al\mbox{.}}{2021}]%
        {fischer2021nekrs}
\bibfield{author}{\bibinfo{person}{Paul Fischer}, \bibinfo{person}{Stefan
  Kerkemeier}, \bibinfo{person}{Misun Min}, \bibinfo{person}{Yu-Hsiang Lan},
  \bibinfo{person}{Malachi Phillips}, \bibinfo{person}{Thilina Rathnayake},
  \bibinfo{person}{Elia Merzari}, \bibinfo{person}{Ananias Tomboulides},
  \bibinfo{person}{Ali Karakus}, \bibinfo{person}{Noel Chalmers}, {and}
  \bibinfo{person}{Tim Warburton}.} \bibinfo{year}{2021}\natexlab{}.
\newblock \bibinfo{title}{NekRS, a GPU-Accelerated Spectral Element
  Navier-Stokes Solver}.
\newblock
\newblock
\showeprint[arxiv]{2104.05829}~[cs.PF]


\bibitem[\protect\citeauthoryear{Fischer, Lottes, and Kerkemeier}{Fischer
  et~al\mbox{.}}{2008}]%
        {fischer2008nek5000}
\bibfield{author}{\bibinfo{person}{Paul~F Fischer}, \bibinfo{person}{James~W
  Lottes}, {and} \bibinfo{person}{Stefan~G Kerkemeier}.}
  \bibinfo{year}{2008}\natexlab{}.
\newblock \bibinfo{title}{nek5000 Web page}.
\newblock
\newblock


\bibitem[\protect\citeauthoryear{Grigoraş, Burovskiy, Luk, and
  Sherwin}{Grigoraş et~al\mbox{.}}{2016}]%
        {7577352}
\bibfield{author}{\bibinfo{person}{Paul Grigoraş}, \bibinfo{person}{Pavel
  Burovskiy}, \bibinfo{person}{Wayne Luk}, {and} \bibinfo{person}{Spencer
  Sherwin}.} \bibinfo{year}{2016}\natexlab{}.
\newblock \showarticletitle{Optimising Sparse Matrix Vector multiplication for
  large scale FEM problems on FPGA}. In \bibinfo{booktitle}{\emph{2016 26th
  International Conference on Field Programmable Logic and Applications
  (FPL)}}. \bibinfo{pages}{1--9}.
\newblock
\urldef\tempurl%
\url{https://doi.org/10.1109/FPL.2016.7577352}
\showDOI{\tempurl}


\bibitem[\protect\citeauthoryear{Gyongyosi and Imre}{Gyongyosi and
  Imre}{2019}]%
        {gyongyosi2019survey}
\bibfield{author}{\bibinfo{person}{Laszlo Gyongyosi} {and}
  \bibinfo{person}{Sandor Imre}.} \bibinfo{year}{2019}\natexlab{}.
\newblock \showarticletitle{A survey on quantum computing technology}.
\newblock \bibinfo{journal}{\emph{Computer Science Review}}
  \bibinfo{volume}{31} (\bibinfo{year}{2019}), \bibinfo{pages}{51--71}.
\newblock


\bibitem[\protect\citeauthoryear{Ivanov, Dryden, Ben-Nun, Li, and
  Hoefler}{Ivanov et~al\mbox{.}}{2021}]%
        {ivanov2021data}
\bibfield{author}{\bibinfo{person}{Andrei Ivanov}, \bibinfo{person}{Nikoli
  Dryden}, \bibinfo{person}{Tal Ben-Nun}, \bibinfo{person}{Shigang Li}, {and}
  \bibinfo{person}{Torsten Hoefler}.} \bibinfo{year}{2021}\natexlab{}.
\newblock \showarticletitle{Data Movement Is All You Need: A Case Study on
  Optimizing Transformers}.
\newblock \bibinfo{journal}{\emph{Proceedings of Machine Learning and Systems}}
   \bibinfo{volume}{3} (\bibinfo{year}{2021}).
\newblock


\bibitem[\protect\citeauthoryear{{Jain}, {Omidian}, {Fraisse}, {Benipal},
  {Liu}, and {Gaitonde}}{{Jain} et~al\mbox{.}}{2020}]%
        {ja-om-20a}
\bibfield{author}{\bibinfo{person}{A.~K. {Jain}}, \bibinfo{person}{H.
  {Omidian}}, \bibinfo{person}{H. {Fraisse}}, \bibinfo{person}{M. {Benipal}},
  \bibinfo{person}{L. {Liu}}, {and} \bibinfo{person}{D. {Gaitonde}}.}
  \bibinfo{year}{2020}\natexlab{}.
\newblock \showarticletitle{A Domain-Specific Architecture for Accelerating
  Sparse Matrix Vector Multiplication on {FPGAs}}. In
  \bibinfo{booktitle}{\emph{Proc. Int. Conf. on Field Programmable Logic and
  Applications (FPL)}}. \bibinfo{pages}{127--132}.
\newblock
\urldef\tempurl%
\url{https://doi.org/10.1109/FPL50879.2020.00031}
\showDOI{\tempurl}


\bibitem[\protect\citeauthoryear{Jansson, Karp, Podobas, Markidis, and
  Schlatter}{Jansson et~al\mbox{.}}{2021}]%
        {jansson2021neko}
\bibfield{author}{\bibinfo{person}{Niclas Jansson}, \bibinfo{person}{Martin
  Karp}, \bibinfo{person}{Artur Podobas}, \bibinfo{person}{Stefano Markidis},
  {and} \bibinfo{person}{Philipp Schlatter}.} \bibinfo{year}{2021}\natexlab{}.
\newblock \bibinfo{title}{Neko: A Modern, Portable, and Scalable Framework for
  High-Fidelity Computational Fluid Dynamics}.
\newblock
\newblock
\showeprint[arxiv]{2107.01243}~[cs.MS]


\bibitem[\protect\citeauthoryear{Jia-Wei and Kung}{Jia-Wei and Kung}{1981}]%
        {jia1981complexity}
\bibfield{author}{\bibinfo{person}{Hong Jia-Wei} {and}
  \bibinfo{person}{Hsiang-Tsung Kung}.} \bibinfo{year}{1981}\natexlab{}.
\newblock \showarticletitle{I/O complexity: The red-blue pebble game}. In
  \bibinfo{booktitle}{\emph{Proceedings of the thirteenth annual ACM symposium
  on Theory of computing}}. \bibinfo{pages}{326--333}.
\newblock


\bibitem[\protect\citeauthoryear{Jouppi, Young, Patil, and Patterson}{Jouppi
  et~al\mbox{.}}{2018}]%
        {jouppi2018motivation}
\bibfield{author}{\bibinfo{person}{Norman Jouppi}, \bibinfo{person}{Cliff
  Young}, \bibinfo{person}{Nishant Patil}, {and} \bibinfo{person}{David
  Patterson}.} \bibinfo{year}{2018}\natexlab{}.
\newblock \showarticletitle{Motivation for and evaluation of the first tensor
  processing unit}.
\newblock \bibinfo{journal}{\emph{IEEE Micro}} \bibinfo{volume}{38},
  \bibinfo{number}{3} (\bibinfo{year}{2018}), \bibinfo{pages}{10--19}.
\newblock


\bibitem[\protect\citeauthoryear{Karp, Podobas, Jansson, Kenter, Plessl,
  Schlatter, and Markidis}{Karp et~al\mbox{.}}{2021}]%
        {karp2021High}
\bibfield{author}{\bibinfo{person}{Martin Karp}, \bibinfo{person}{Artur
  Podobas}, \bibinfo{person}{Niclas Jansson}, \bibinfo{person}{Tobias Kenter},
  \bibinfo{person}{Christian Plessl}, \bibinfo{person}{Philipp Schlatter},
  {and} \bibinfo{person}{Stefano Markidis}.} \bibinfo{year}{2021}\natexlab{}.
\newblock \showarticletitle{High-Performance Spectral Element Methods on
  Field-Programmable Gate Arrays : Implementation, Evaluation, and Future
  Projection}. In \bibinfo{booktitle}{\emph{2021 IEEE International Parallel
  and Distributed Processing Symposium (IPDPS)}}. \bibinfo{pages}{1077--1086}.
\newblock
\urldef\tempurl%
\url{https://doi.org/10.1109/IPDPS49936.2021.00116}
\showDOI{\tempurl}


\bibitem[\protect\citeauthoryear{Karypis, Schloegel, and Kumar}{Karypis
  et~al\mbox{.}}{1997}]%
        {karypis1997parmetis}
\bibfield{author}{\bibinfo{person}{George Karypis}, \bibinfo{person}{Kirk
  Schloegel}, {and} \bibinfo{person}{Vipin Kumar}.}
  \bibinfo{year}{1997}\natexlab{}.
\newblock \showarticletitle{Parmetis: Parallel graph partitioning and sparse
  matrix ordering library}.
\newblock  (\bibinfo{year}{1997}).
\newblock


\bibitem[\protect\citeauthoryear{Kenter}{Kenter}{2019}]%
        {ke-19a}
\bibfield{author}{\bibinfo{person}{Tobias Kenter}.}
  \bibinfo{year}{2019}\natexlab{}.
\newblock \showarticletitle{Invited Tutorial: {OpenCL} design flows for {Intel}
  and {Xilinx} {FPGAs}: Using common design patterns and dealing with
  vendor-specific differences}. In \bibinfo{booktitle}{\emph{Proc. Int.
  Workshop on FPGAs for Software Programmers (FSP), collocated with Int. Conf.
  on Field Programmable Logic and Applications (FPL)}}.
\newblock


\bibitem[\protect\citeauthoryear{Kenter, Mahale, Alhaddad, Grynko, Schmitt,
  Afzal, Hannig, Förstner, and Plessl}{Kenter et~al\mbox{.}}{2018}]%
        {8457652}
\bibfield{author}{\bibinfo{person}{Tobias Kenter}, \bibinfo{person}{Gopinath
  Mahale}, \bibinfo{person}{Samer Alhaddad}, \bibinfo{person}{Yevgen Grynko},
  \bibinfo{person}{Christian Schmitt}, \bibinfo{person}{Ayesha Afzal},
  \bibinfo{person}{Frank Hannig}, \bibinfo{person}{Jens Förstner}, {and}
  \bibinfo{person}{Christian Plessl}.} \bibinfo{year}{2018}\natexlab{}.
\newblock \showarticletitle{OpenCL-Based FPGA Design to Accelerate the Nodal
  Discontinuous Galerkin Method for Unstructured Meshes}. In
  \bibinfo{booktitle}{\emph{2018 IEEE 26th Annual International Symposium on
  Field-Programmable Custom Computing Machines (FCCM)}}.
  \bibinfo{pages}{189--196}.
\newblock
\urldef\tempurl%
\url{https://doi.org/10.1109/FCCM.2018.00037}
\showDOI{\tempurl}


\bibitem[\protect\citeauthoryear{Kenter, Shambhu, Faghih-Naini, and
  Aizinger}{Kenter et~al\mbox{.}}{2021}]%
        {ke-sh-21i}
\bibfield{author}{\bibinfo{person}{Tobias Kenter}, \bibinfo{person}{Adesh
  Shambhu}, \bibinfo{person}{Sara Faghih-Naini}, {and} \bibinfo{person}{Vadym
  Aizinger}.} \bibinfo{year}{2021}\natexlab{}.
\newblock \showarticletitle{Algorithm-Hardware Co-design of a Discontinuous
  {Galerkin} Shallow-Water Model for a Dataflow Architecture on {FPGA}}. In
  \bibinfo{booktitle}{\emph{Proc. Platform for Advanced Scientific Computing
  Conf. (PASC)}}.
\newblock
\newblock
\shownote{To appear.}


\bibitem[\protect\citeauthoryear{Kuon, Tessier, and Rose}{Kuon
  et~al\mbox{.}}{2008}]%
        {kuon2008fpga}
\bibfield{author}{\bibinfo{person}{Ian Kuon}, \bibinfo{person}{Russell
  Tessier}, {and} \bibinfo{person}{Jonathan Rose}.}
  \bibinfo{year}{2008}\natexlab{}.
\newblock \bibinfo{booktitle}{\emph{FPGA architecture: Survey and challenges}}.
\newblock \bibinfo{publisher}{Now Publishers Inc}.
\newblock


\bibitem[\protect\citeauthoryear{Kwasniewski, Ben-Nun, Gianinazzi, Calotoiu,
  Schneider, Ziogas, Besta, and Hoefler}{Kwasniewski et~al\mbox{.}}{2021}]%
        {Kwasniewski_2021}
\bibfield{author}{\bibinfo{person}{Grzegorz Kwasniewski}, \bibinfo{person}{Tal
  Ben-Nun}, \bibinfo{person}{Lukas Gianinazzi}, \bibinfo{person}{Alexandru
  Calotoiu}, \bibinfo{person}{Timo Schneider},
  \bibinfo{person}{Alexandros~Nikolaos Ziogas}, \bibinfo{person}{Maciej Besta},
  {and} \bibinfo{person}{Torsten Hoefler}.} \bibinfo{year}{2021}\natexlab{}.
\newblock \showarticletitle{Pebbles, Graphs, and a Pinch of Combinatorics}.
\newblock \bibinfo{journal}{\emph{Proceedings of the 33rd ACM Symposium on
  Parallelism in Algorithms and Architectures}} (\bibinfo{date}{Jul}
  \bibinfo{year}{2021}).
\newblock
\showISBNx{9781450380706}
\urldef\tempurl%
\url{https://doi.org/10.1145/3409964.3461796}
\showDOI{\tempurl}


\bibitem[\protect\citeauthoryear{Kwasniewski, Kabi{\'c}, Besta, VandeVondele,
  Solc{\`a}, and Hoefler}{Kwasniewski et~al\mbox{.}}{2019}]%
        {kwasniewski2019red}
\bibfield{author}{\bibinfo{person}{Grzegorz Kwasniewski},
  \bibinfo{person}{Marko Kabi{\'c}}, \bibinfo{person}{Maciej Besta},
  \bibinfo{person}{Joost VandeVondele}, \bibinfo{person}{Raffaele Solc{\`a}},
  {and} \bibinfo{person}{Torsten Hoefler}.} \bibinfo{year}{2019}\natexlab{}.
\newblock \showarticletitle{Red-blue pebbling revisited: near optimal parallel
  matrix-matrix multiplication}. In \bibinfo{booktitle}{\emph{Proceedings of
  the International Conference for High Performance Computing, Networking,
  Storage and Analysis}}. \bibinfo{pages}{1--22}.
\newblock


\bibitem[\protect\citeauthoryear{Marjanovi{\'c}, Gracia, and
  Glass}{Marjanovi{\'c} et~al\mbox{.}}{2014}]%
        {marjanovic2014performance}
\bibfield{author}{\bibinfo{person}{Vladimir Marjanovi{\'c}},
  \bibinfo{person}{Jos{\'e} Gracia}, {and} \bibinfo{person}{Colin~W Glass}.}
  \bibinfo{year}{2014}\natexlab{}.
\newblock \showarticletitle{Performance modeling of the HPCG benchmark}. In
  \bibinfo{booktitle}{\emph{International Workshop on Performance Modeling,
  Benchmarking and Simulation of High Performance Computer Systems}}. Springer,
  \bibinfo{pages}{172--192}.
\newblock


\bibitem[\protect\citeauthoryear{Meyer, Kenter, and Plessl}{Meyer
  et~al\mbox{.}}{2020}]%
        {me-ke-20a}
\bibfield{author}{\bibinfo{person}{Marius Meyer}, \bibinfo{person}{Tobias
  Kenter}, {and} \bibinfo{person}{Christian Plessl}.}
  \bibinfo{year}{2020}\natexlab{}.
\newblock \showarticletitle{Evaluating {FPGA} Accelerator Performance with a
  Parameterized OpenCL Adaptation of Selected Benchmarks of the HPCChallenge
  Benchmark Suite}. In \bibinfo{booktitle}{\emph{Proc. Workshop on
  Heterogeneous High-performance Reconfigurable Computing (H2RC), held in
  conjuction with Int. Conf. on High Performance Computing, Networking, Storage
  and Analysis (SC)}}. \bibinfo{pages}{10--18}.
\newblock
\urldef\tempurl%
\url{https://doi.org/10.1109/H2RC51942.2020.00007}
\showDOI{\tempurl}


\bibitem[\protect\citeauthoryear{Nagy, Nemes, Hiba, Kiss, Cs{\'\i}k, and
  Szolgay}{Nagy et~al\mbox{.}}{2012}]%
        {nagy2012fpga}
\bibfield{author}{\bibinfo{person}{Zolt{\'a}n Nagy}, \bibinfo{person}{Csaba
  Nemes}, \bibinfo{person}{Antal Hiba}, \bibinfo{person}{Andr{\'a}s Kiss},
  \bibinfo{person}{{\'A}rp{\'a}d Cs{\'\i}k}, {and} \bibinfo{person}{P{\'e}ter
  Szolgay}.} \bibinfo{year}{2012}\natexlab{}.
\newblock \showarticletitle{FPGA based acceleration of computational fluid flow
  simulation on unstructured mesh geometry}. In \bibinfo{booktitle}{\emph{22nd
  International Conference on Field Programmable Logic and Applications
  (FPL)}}. IEEE, \bibinfo{pages}{128--135}.
\newblock


\bibitem[\protect\citeauthoryear{Papakonstantinou, Gururaj, Stratton, Chen,
  Cong, and Hwu}{Papakonstantinou et~al\mbox{.}}{2009}]%
        {papakonstantinou2009fcuda}
\bibfield{author}{\bibinfo{person}{Alexandros Papakonstantinou},
  \bibinfo{person}{Karthik Gururaj}, \bibinfo{person}{John~A Stratton},
  \bibinfo{person}{Deming Chen}, \bibinfo{person}{Jason Cong}, {and}
  \bibinfo{person}{Wen-Mei~W Hwu}.} \bibinfo{year}{2009}\natexlab{}.
\newblock \showarticletitle{FCUDA: Enabling efficient compilation of CUDA
  kernels onto FPGAs}. In \bibinfo{booktitle}{\emph{2009 IEEE 7th Symposium on
  Application Specific Processors}}. IEEE, \bibinfo{pages}{35--42}.
\newblock


\bibitem[\protect\citeauthoryear{Pilato and Ferrandi}{Pilato and
  Ferrandi}{2013}]%
        {pilato2013bambu}
\bibfield{author}{\bibinfo{person}{Christian Pilato} {and}
  \bibinfo{person}{Fabrizio Ferrandi}.} \bibinfo{year}{2013}\natexlab{}.
\newblock \showarticletitle{Bambu: A modular framework for the high level
  synthesis of memory-intensive applications}. In
  \bibinfo{booktitle}{\emph{2013 23rd International Conference on Field
  programmable Logic and Applications}}. IEEE, \bibinfo{pages}{1--4}.
\newblock


\bibitem[\protect\citeauthoryear{Podobas and Brorsson}{Podobas and
  Brorsson}{2016}]%
        {podobas2016empowering}
\bibfield{author}{\bibinfo{person}{Artur Podobas} {and} \bibinfo{person}{Mats
  Brorsson}.} \bibinfo{year}{2016}\natexlab{}.
\newblock \showarticletitle{Empowering openmp with automatically generated
  hardware}. In \bibinfo{booktitle}{\emph{2016 International Conference on
  Embedded Computer Systems: Architectures, Modeling and Simulation (SAMOS)}}.
  IEEE, \bibinfo{pages}{245--252}.
\newblock


\bibitem[\protect\citeauthoryear{Podobas and Matsuoka}{Podobas and
  Matsuoka}{2017}]%
        {podobas2017designing}
\bibfield{author}{\bibinfo{person}{Artur Podobas} {and}
  \bibinfo{person}{Satoshi Matsuoka}.} \bibinfo{year}{2017}\natexlab{}.
\newblock \showarticletitle{Designing and accelerating spiking neural networks
  using OpenCL for FPGAs}. In \bibinfo{booktitle}{\emph{2017 International
  Conference on Field Programmable Technology (ICFPT)}}. IEEE,
  \bibinfo{pages}{255--258}.
\newblock


\bibitem[\protect\citeauthoryear{Podobas, Sano, and Matsuoka}{Podobas
  et~al\mbox{.}}{2020a}]%
        {podobas2020survey}
\bibfield{author}{\bibinfo{person}{Artur Podobas}, \bibinfo{person}{Kentaro
  Sano}, {and} \bibinfo{person}{Satoshi Matsuoka}.}
  \bibinfo{year}{2020}\natexlab{a}.
\newblock \showarticletitle{A survey on coarse-grained reconfigurable
  architectures from a performance perspective}.
\newblock \bibinfo{journal}{\emph{IEEE Access}}  \bibinfo{volume}{8}
  (\bibinfo{year}{2020}), \bibinfo{pages}{146719--146743}.
\newblock


\bibitem[\protect\citeauthoryear{Podobas, Sano, and Matsuoka}{Podobas
  et~al\mbox{.}}{2020b}]%
        {podobas2020template}
\bibfield{author}{\bibinfo{person}{Artur Podobas}, \bibinfo{person}{Kentaro
  Sano}, {and} \bibinfo{person}{Satoshi Matsuoka}.}
  \bibinfo{year}{2020}\natexlab{b}.
\newblock \showarticletitle{A template-based framework for exploring
  coarse-grained reconfigurable architectures}. In
  \bibinfo{booktitle}{\emph{2020 IEEE 31st International Conference on
  Application-specific Systems, Architectures and Processors (ASAP)}}. IEEE,
  \bibinfo{pages}{1--8}.
\newblock


\bibitem[\protect\citeauthoryear{Qasaimeh, Denolf, Lo, Vissers, Zambreno, and
  Jones}{Qasaimeh et~al\mbox{.}}{2019}]%
        {qasaimeh2019comparing}
\bibfield{author}{\bibinfo{person}{Murad Qasaimeh}, \bibinfo{person}{Kristof
  Denolf}, \bibinfo{person}{Jack Lo}, \bibinfo{person}{Kees Vissers},
  \bibinfo{person}{Joseph Zambreno}, {and} \bibinfo{person}{Phillip~H Jones}.}
  \bibinfo{year}{2019}\natexlab{}.
\newblock \showarticletitle{Comparing energy efficiency of CPU, GPU and FPGA
  implementations for vision kernels}. In \bibinfo{booktitle}{\emph{2019 IEEE
  international conference on embedded software and systems (ICESS)}}. IEEE,
  \bibinfo{pages}{1--8}.
\newblock


\bibitem[\protect\citeauthoryear{Sano}{Sano}{2013}]%
        {sano2013fpga}
\bibfield{author}{\bibinfo{person}{Kentaro Sano}.}
  \bibinfo{year}{2013}\natexlab{}.
\newblock \showarticletitle{FPGA-based systolic computational-memory array for
  scalable stencil computations}.
\newblock In \bibinfo{booktitle}{\emph{High-Performance Computing Using
  FPGAs}}. \bibinfo{publisher}{Springer}, \bibinfo{pages}{279--303}.
\newblock


\bibitem[\protect\citeauthoryear{Schuman, Potok, Patton, Birdwell, Dean, Rose,
  and Plank}{Schuman et~al\mbox{.}}{2017}]%
        {schuman2017survey}
\bibfield{author}{\bibinfo{person}{Catherine~D Schuman},
  \bibinfo{person}{Thomas~E Potok}, \bibinfo{person}{Robert~M Patton},
  \bibinfo{person}{J~Douglas Birdwell}, \bibinfo{person}{Mark~E Dean},
  \bibinfo{person}{Garrett~S Rose}, {and} \bibinfo{person}{James~S Plank}.}
  \bibinfo{year}{2017}\natexlab{}.
\newblock \showarticletitle{A survey of neuromorphic computing and neural
  networks in hardware}.
\newblock \bibinfo{journal}{\emph{arXiv preprint arXiv:1705.06963}}
  (\bibinfo{year}{2017}).
\newblock


\bibitem[\protect\citeauthoryear{Slotnick, Khodadoust, Alonso, Darmofal, Gropp,
  Lurie, and Mavriplis}{Slotnick et~al\mbox{.}}{2014}]%
        {slotnick2014cfd}
\bibfield{author}{\bibinfo{person}{Jeffrey~P Slotnick},
  \bibinfo{person}{Abdollah Khodadoust}, \bibinfo{person}{Juan Alonso},
  \bibinfo{person}{David Darmofal}, \bibinfo{person}{William Gropp},
  \bibinfo{person}{Elizabeth Lurie}, {and} \bibinfo{person}{Dimitri~J
  Mavriplis}.} \bibinfo{year}{2014}\natexlab{}.
\newblock \showarticletitle{CFD vision 2030 study: a path to revolutionary
  computational aerosciences}.
\newblock  (\bibinfo{year}{2014}).
\newblock


\bibitem[\protect\citeauthoryear{Vetter, DeBenedictis, and Conte}{Vetter
  et~al\mbox{.}}{2017}]%
        {vetter2017architectures}
\bibfield{author}{\bibinfo{person}{Jeffrey~S Vetter}, \bibinfo{person}{Erik~P
  DeBenedictis}, {and} \bibinfo{person}{Thomas~M Conte}.}
  \bibinfo{year}{2017}\natexlab{}.
\newblock \showarticletitle{Architectures for the post-Moore era}.
\newblock \bibinfo{journal}{\emph{IEEE Micro}} \bibinfo{volume}{37},
  \bibinfo{number}{04} (\bibinfo{year}{2017}), \bibinfo{pages}{6--8}.
\newblock


\bibitem[\protect\citeauthoryear{Waldrop}{Waldrop}{2016}]%
        {waldrop2016chips}
\bibfield{author}{\bibinfo{person}{M~Mitchell Waldrop}.}
  \bibinfo{year}{2016}\natexlab{}.
\newblock \showarticletitle{The chips are down for Moore’s law}.
\newblock \bibinfo{journal}{\emph{Nature News}} \bibinfo{volume}{530},
  \bibinfo{number}{7589} (\bibinfo{year}{2016}), \bibinfo{pages}{144}.
\newblock


\bibitem[\protect\citeauthoryear{Zhang, Li, Lu, Jantsch, Gao, Pan, and
  Han}{Zhang et~al\mbox{.}}{2014}]%
        {zhang2014survey}
\bibfield{author}{\bibinfo{person}{Yuang Zhang}, \bibinfo{person}{Li Li},
  \bibinfo{person}{Zhonghai Lu}, \bibinfo{person}{Axel Jantsch},
  \bibinfo{person}{Minglun Gao}, \bibinfo{person}{Hongbing Pan}, {and}
  \bibinfo{person}{Feng Han}.} \bibinfo{year}{2014}\natexlab{}.
\newblock \showarticletitle{A survey of memory architecture for 3D chip
  multi-processors}.
\newblock \bibinfo{journal}{\emph{Microprocessors and Microsystems}}
  \bibinfo{volume}{38}, \bibinfo{number}{5} (\bibinfo{year}{2014}),
  \bibinfo{pages}{415--430}.
\newblock


\bibitem[\protect\citeauthoryear{Zohouri and Matsuoka}{Zohouri and
  Matsuoka}{2019}]%
        {8945518}
\bibfield{author}{\bibinfo{person}{Hamid~Reza Zohouri} {and}
  \bibinfo{person}{Satoshi Matsuoka}.} \bibinfo{year}{2019}\natexlab{}.
\newblock \showarticletitle{The Memory Controller Wall: Benchmarking the Intel
  FPGA SDK for OpenCL Memory Interface}. In \bibinfo{booktitle}{\emph{2019
  IEEE/ACM International Workshop on Heterogeneous High-performance
  Reconfigurable Computing (H2RC)}}. \bibinfo{pages}{11--18}.
\newblock
\urldef\tempurl%
\url{https://doi.org/10.1109/H2RC49586.2019.00007}
\showDOI{\tempurl}


\bibitem[\protect\citeauthoryear{Zohouri, Podobas, and Matsuoka}{Zohouri
  et~al\mbox{.}}{2018}]%
        {zohouri2018combined}
\bibfield{author}{\bibinfo{person}{Hamid~Reza Zohouri}, \bibinfo{person}{Artur
  Podobas}, {and} \bibinfo{person}{Satoshi Matsuoka}.}
  \bibinfo{year}{2018}\natexlab{}.
\newblock \showarticletitle{Combined spatial and temporal blocking for
  high-performance stencil computation on FPGAs using OpenCL}. In
  \bibinfo{booktitle}{\emph{Proceedings of the 2018 ACM/SIGDA International
  Symposium on Field-Programmable Gate Arrays}}. \bibinfo{pages}{153--162}.
\newblock


\end{thebibliography}

\end{document}